%% file: tark98rjcorr.tex
\documentstyle[12pt,chicagob]{article}

\input bookdefn
\newcommand{\CorrectB}{\tilde{B}}
\newcommand{\Natp}{{\bf N}^+}
\newcommand{\doact}{{\sl do}}
\renewcommand{\next}{\Circ}
\newcommand{\intension}[1]{[\![ #1 ]\!]}
\newcommand{\mini}{{\sf min}_i}

\newcommand{\ogen}{o}
\newcommand{\rgen}{\sigma}
\newcommand{\rgend}{\sigma^*}

\newcommand{\cJ}{{\cal J}}
\newcommand{\Jc}{\cJ^c}
\newcommand{\Jp}{\cJ^+}
\newcommand{\rank}{\kappa}
\renewcommand{\L}{{\cal L}}
\newcommand{\Lb}{{\L_B}}
\newcommand{\Lbc}{{\L^\RCond_B}}
\newcommand{\Lc}{{\L^\RCond}}
\newcommand{\Lk}{{\L_K}}

\newcommand{\lta}{\mbox{{$<${\hskip -6.8pt}$<$}}}
\newcommand{\closest}{{\tt closest}}
\newcommand{\close}{{\tt close}}
\newcommand{\latest}{{\it last}}
\newcommand{\RCond}{>}

\newcommand{\Rrep}{{\bf R}}
\newcommand{\Irep}{{\bf I}}
\newcommand{\Ip}{{\cal I}^+}
\newcommand{\Rp}{\cR^+}
\newcommand{\PgbtR}{\Pgbt^{{\scriptscriptstyle >}}}
\newcommand{\PgbtK}{\Pgbt^{K}}

\newcommand{\PgbtB}{\Pgbt^{\Diamond B}}

\renewcommand{\Pgkb}{{\sf Pg}_{{\it kb}}}

\newcommand{\Pgcb}{{\sf Pg}_{{\it cb}}}
\newcommand{\Pkb}{\Pgkb}

\newcommand{\noop}{{\sf skip}}
\newcommand{\NOOP}{{\sf SKIP}}

\renewcommand{\clock}{\mbox{\it time}}
\newcommand{\bruns}{\mbox{{\sf b}-runs}}
\newcommand{\bpoints}{\mbox{{\sf b}-pts}}
\newcommand{\Po}{P^\omega}
\newcommand{\Ro}{R^\omega}
\newcommand{\cJo}{\cJ^\omega}

\begin{document}
\begin{titlepage}
\title{Using Counterfactuals in Knowledge-Based Programming}
\author{
Joseph Y.\ Halpern%
\thanks{The work was supported in part by NSF under
grant IRI-96-25901 and IIS--0090145, by the Air Force Office of
Scientific Research under grant F49620-96-1-0323, and by ONR under
grants N00014-00-1-03-41 and N00014-01-1-0795.
A preliminary version of this paper appeared in the Proceedings of the
Seventh Conference on Theoretical Aspects of Rationality and Knowledge
(TARK), 1998.}
\\
Cornell University\\
Dept. of Computer Science\\
Ithaca, NY 14853\\
halpern@cs.cornell.edu\\
http://www.cs.cornell.edu/home/halpern
\and
Yoram Moses\\
Department of Electrical Engineering\\
Technion---Israel Institute of Technology\\
32000 Haifa, Israel\\
moses@ee.technion.ac.il}

\date{\today}
\maketitle
\thispagestyle{empty}

\begin{abstract}
This paper adds counterfactuals to the framework of
{\em knowledge-based programs\/}
of Fagin, Halpern, Moses, and Vardi \citeyear{FHMV,FHMV94}. 
The use of counterfactuals is illustrated by designing 
a protocol in which an agent stops sending messages 
once it knows that it is safe to do so.
Such behavior is difficult to capture in the original 
framework because it involves reasoning about counterfactual 
executions, including ones that are not consistent with 
the protocol. 
Attempts to formalize these notions without counterfactuals 
are shown to lead to 
rather counterintuitive behavior.  
\end{abstract}
\end{titlepage}

\section{Introduction}
{\em Knowledge-based programs}, first introduced 
by Halpern and Fagin \citeyear{HF87} and
further developed by Fagin, Halpern, Moses, and Vardi \citeyear{FHMV,FHMV94}, 
are intended to provide a high-level framework for the design
and specification of protocols.  The idea is that, in knowledge-based
programs, there are explicit tests for knowledge.  Thus, a
knowledge-based program might have the form
$$\begin{array}{l}
\mbox{{\bf if} $K(x=0)$ {\bf then} $y := y+1$ {\bf else} $\noop$},\\
\end{array}$$
where $K(x=0)$ should be read as ``you know $x=0$''
and $\noop$ is the action of doing nothing.
We can informally view this knowledge-based program
as saying ``if you know that $x=0$, then set $y$ to $y+1$
(otherwise do nothing)''.

Knowledge-based programs are an attempt to capture the intuition that
what an agent does depends on what it knows.  
They have been used successfully 
in papers such as \cite{DM,Had,HMW,HZ,Maz,Maz90,MT,NT} both to
help in the design of new protocols and to clarify the understanding of
existing protocols.  However, as we show here, there are cases when,
used naively,
knowledge-based programs exhibit some quite counterintuitive behavior.
We 
then 
show how this can be overcome by the use of
{\em counterfactuals\/} \cite{Lewis73,Stalnaker68}.
In this introduction, we discuss these issues informally, leaving the
formal details to later sections of the paper.

Some counterintuitive aspects of knowledge-based programs can be 
understood by considering the {\em bit-transmission problem\/} from
\cite{FHMV}.
In this problem, there are two processes, a
{\em sender\/}~$S$ and a {\em receiver}~$R$, that communicate over
a communication line.  The sender starts with one bit (either 0 or 1)
that it wants to communicate to the receiver.
The communication line may be
faulty and lose messages in either direction in any given round.
That is, there is no guarantee that
a message sent by either~$S$ or~$R$ will be received.
Because of the uncertainty regarding possible message loss,
$S$ sends the bit to~$R$
in every round, until~$S$ receives an $\ack$ message
from~$R$ acknowledging receipt of the bit.
$R$ starts sending the $\ack$ message
in the round after it receives the bit, and continues to send it
repeatedly from then on.
The sender $S$ can be viewed as 
running
the program $\Pgbt_S$:
$$\mbox{{\bf if} $\recack$ {\bf then} \noop~{\bf else}
{\sf sendbit},}$$
where $\recack$ is a proposition that is true if $S$ has 
already received an $\ack$ message from~$R$ and false otherwise, while 
{\sf sendbit} is the action of sending the bit.%
\footnote{Running such a program amounts to performing the statement repeatedly
forever.}
Note that $\Pgbt_S$ is a {\em standard\/} program---it
does not have tests for knowledge.  We can capture some of the
intuitions behind this program by using knowledge.
The sender~$S$ keeps sending the bit until an
acknowledgment is received from the receiver~$R$.
Thus, another way to describe the sender's behavior is to say that~$S$ 
keeps sending the bit until it {\em knows\/} that the bit was
received by~$R$.
This behavior can be characterized by the knowledge-based program~$\Pgbt'_S$:
$$\mbox{{\bf if} $K_S(\recbit)$ {\bf then} \noop~{\bf else}
{\sf sendbit},}$$
where $\recbit$ is a proposition that is true once $R$ has received the
bit.
The advantage of this program over the standard
program $\Pgbt_S$ is that it abstracts away
the mechanism by which~$S$ learns that the bit was received by~$R$.
For example, if messages from~$S$ to~$R$ are guaranteed to be
delivered in the same round in which they are sent, then~$S$ 
knows that~$R$ received the bit even if~$S$ does not receive an 
acknowledgment.

We might hope to improve this even further.  
Consider a system where all messages sent are guaranteed to be
delivered, but rather than arriving in one round, they spend exactly
five rounds in transit. 
In such a system, 
a sender using~$\Pgbt_S$ will send the bit 10 times, because it will 
take 10 rounds to get the receiver's
acknowledgment after the original message is sent.  
The program~$\Pgbt'_S$ is somewhat better; using it~$S$
sends the bit only five times, since after the fifth round, $S$ will 
know that $R$ got his first message.  Nevertheless, this seems wasteful.  
Given that messages are guaranteed to be delivered, it clearly suffices for the
sender to send the bit once.  Intuitively, the sender should be
able to stop sending the message as soon as it knows that the receiver
will {\em eventually\/} receive 
a copy of 
the message; 
the sender 
should not have to wait until the receiver 
{\em actually\/} receives it.

It seems that there should be 
no problem handling this using knowledge-based programs.  
Let $\Diamond$ be the standard ``eventually'' operator from temporal 
logic \cite{MP1}; $\Diamond \phi$ means that $\phi$ is eventually true, 
and let $\Box$ be its dual, ``always''.
Now the following knowledge-based program $\Pgbt^*_S$ for the sender
should
capture exactly what is required:
$$\mbox{{\bf if} $K_S (\Diamond \recbit)$ {\bf then} \noop~{\bf else}
{\sf sendbit}.}$$

Unfortunately, $\Pgbt^*_S$  does not capture our intuitions here.
To understand why, consider the sender $S$. 
Should it send the bit in the first round?
According to $\Pgbt^*_S$, the sender $S$
should send the bit if $S$ does not know that $R$ will eventually
receive the bit.
But if $S$ sends the bit, then $S$ knows that $R$ will
eventually receive it (since messages are guaranteed to be delivered in
5 rounds).  Thus, $S$ should not send the bit.  Similar
arguments show that $S$ should not send the bit at any round.  On the
other hand, if $S$ never sends the bit,
then $R$ will never receive it and thus $S$ 
{\em should} send the bit!
It follows that according to $\Pgbt^*_S$, 
$S$ should send the bit exactly if it will never send the bit. 
Obviously, there is no way~$S$ can follow such a program. 
Put another way, this program cannot be implemented by a standard 
program at all.  
This is
certainly not the behavior we would intuitively have expected of
$\Pgbt_S^*$.%
\footnote{While intuitions may, of course, vary,
some evidence of the counterintuitive behavior of this program is that
it was used in a draft of \cite{FHMV}; it was several months before we
realized its problematic nature.}

One approach to dealing with this problem is to change the semantics of
knowledge-based programs. 
Inherent in the semantics of knowledge-based programs 
is the fact that an agent knows what standard protocol she is
following. Thus, if the sender is guaranteed to send a message in round~two, 
then she knows at time one that the message will be sent in the following
round. Moreover, if communication is reliable, she also knows the message 
will later be received. 
If we weaken the semantics of knowledge sufficiently,
then this problem disappears. 
(See \cite{EMM98} for an approach to dealing with 
the problem addressed in this paper 
along these lines.)  However, it is not yet clear 
how to 
make this change and still maintain the attractive features of
knowledge-based programs that we discussed earlier.

In this paper we consider another approach to dealing with 
the
problem, based on counterfactuals. 
Our claim is that the program $\Pgbt^*_S$ does not
adequately capture our intuitions.  Rather than saying that~$S$
should stop sending if~$S$ knows that~$R$ will eventually receive
the bit we should, instead, say that~$S$ should stop sending if it
knows that {\em even if~$S$ does not send another message\/} $R$ will
eventually receive the bit.

How should we capture this?
Let $\doact(i,{\sf a})$ be the formula 
that is true at a point $(r,m)$ if process $i$ performs
${\sf a}$ in the next round.%
\footnote{We assume that round $m$ takes place between time $m-1$ and
$m$.  Thus, the next round after $(r,m)$ is round $m+1$, which takes takes
place between $(r,m)$ and $(r,m+1)$.}  
The most obvious way to
capture ``(even) if $S$ does not send a message then $R$ will eventually 
receive the bit'' uses 
standard implication,
also known as {\em material implication\/} or {\em material
conditional\/} in philosophical logic:
$\doact(S,\noop) \rimp \recbit$.  This leads to a program such as
$\Pgbt_S^{\rimp}$:
$$\mbox{{\bf if} $K_S (\doact(S,\noop) \rimp \Diamond \recbit)$
{\bf then} \noop~{\bf else} {\sf sendbit}.}$$

Unfortunately, this program does not solve our problems.  
It, too is not implementable by a standard program. 
To see why, suppose
that
there is some point in the execution of this protocol where $S$
sends a message.  At this point~$S$ knows it is sending a message, 
so $S$ knows that $\doact(S,\noop)$ is false.  Thus, $S$ knows that
$\doact(S,\noop) \rimp \Diamond \recbit$ holds. As a result, 
$K_S(\doact(S,\noop) \rimp \Diamond \recbit)$ is true, 
so that the test 
in~$\Pgbt_S^{\rimp}$ succeeds. 
Thus, 
according to~$\Pgbt_S^{\rimp}$, the sender~$S$ 
should {\em not} send a message 
at this point.
On the other hand, if $S$ {\em never\/} sends a message according to the 
protocol (under any circumstance), then $S$ knows that it will never send 
a message (since, after all, $S$ knows how the protocol works).  
But in this case, $S$ knows that the receiver will never receive the
bit, so the test 
fails.
Thus, 
according to~$\Pgbt_S^{\rimp}$, the sender~$S$ should send the message  
as its first action, this time contradicting the assumption that the message 
is never sent.
Nothing that $S$ can do is consistent with this program.

The problem here is the use of 
material implication ($\rimp$).
Our intuitions are better captured by using 
counterfactual implication, which we denote by~$\RCond$.
A statement such as
$\phi \RCond \psi$ is read ``if $\phi$ then $\psi$'', just
like $\phi \rimp \psi$.
However, the semantics
 of $\RCond$ is very different from that of~$\rimp$.  
The idea, which goes back to  Stalnaker
\citeyear{Stalnaker68} and Lewis \citeyear{Lewis73} 
is that a statement such as $\phi \RCond \psi$ is true at a world $w$ if in the
worlds ``closest to'' or ``most like'' $w$ where $\phi$ is true, $\psi$
is also true. 
This attempts to capture the intuition that the counterfactual statement 
$\phi \RCond \psi$ stands for ``if $\phi$ were the case, 
then $\psi$ would hold''. 
For example, suppose 
that
we have
a wet match and we make a statement such as ``if the match were dry then
it would light''.  Using $\rimp$ this statement is trivially true, since
the antecedent is false.  However, with $\RCond$, the situation is not
so obvious.  We must consider the worlds most like the actual world
where the match is in fact dry and decide whether it would light in
those worlds.  If we think the match is defective for some reason, then
even if it were dry, it would not light.

A central issue in the application of counterfactual reasoning to a concrete 
problem is that we 
need to specify what the ``closest worlds'' are.
The philosophical literature does not give us any guidance on this point.
We present some general approaches for doing so, motivated by our
interest in modeling counterfactual reasoning about what would happen 
if an agent were to deviate from the protocol it is following. 
We believe that this example can inform similar applications of 
counterfactual reasoning in other contexts. 
There is a subtle technical point 
that needs to be addressed in order to use counterfactuals 
in knowledge-based programs.
Traditionally, we talk about a knowledge-based program~$\Pgkb$ being 
implemented by a protocol~$P$. This is the case when the behavior
prescribed by~$P$ is in accordance with what~$\Pgkb$ specifies. 
To determine whether~$P$ implements~$\Pgkb$, the knowledge tests
(tests for the truth of formulas of the form $K_i\phi$) in~$\Pgkb$ are 
evaluated with respect to the points appearing in the set of runs of~$P$. 
In this system, all the agents know that the properties of $P$
(e.g. facts like process 1 always sending an acknowledgment after
receiving a message from process 2) hold in all runs.
But this set of runs does not account for what may happen if 
(counter to fact) some agents were to deviate 
from~$P$. 
In counterfactual reasoning, we need to evaluate 
formulas with respect to a larger set of 
runs that allows for such deviations. 

We deal with this problem by evaluating counterfactuals with respect
to a system consisting of all possible runs (not just the ones generated
by $P$).  
While working with this larger system enables us to reason about
counterfactuals, processes no longer know the properties of $P$ in this
system, since it includes many runs not in $P$.
In order to deal with this, we add a notion of likelihood to the system
using what are called {\em ranking functions\/} \cite{spohn:88}.
Runs generated by $P$ get rank 0; all other runs get higher rank.
(Lower ranks imply greater likelihood.)  Ranks let us define a standard
notion of {\em belief}.  Although a process does not {\em know\/} that
the properties of $P$ hold, it {\em believes\/} that they do.
Moreover, when
restricted to the set of runs of the original protocol~$P$, this 
notion of belief satisfies the knowledge axiom $B_i\phi\rimp\phi$, 
and coincides with the notion of knowledge we had in the original system. 
Thus, when the original protocol is followed, our notion of 
belief acts essentially like knowledge. 

Using the counterfactual operator and this interpretation for belief, 
we get the program 
$\PgbtR_S$:
$$\mbox{{\bf if} $B_S (\doact(S,\noop) \RCond \Diamond \recbit)$
{\bf then} \noop~{\bf else} {\sf sendbit}.}$$
We show that using counterfactuals 
in this way has the desired effect here.  
If message delivery is guaranteed, then after the 
message has been sent once, under what seems to be the most reasonable
interpretation of ``the closest world'' where the message is not sent,
the sender believes that the bit will eventually be received.
In particular, in contexts where messages are delivered in
five rounds, using $\PgbtR_S$, the sender will send one message.

As we said, one advantage of $\Pgbt_S'$ over the standard
program $\Pgbt_S$ is that it abstracts away
the mechanism by which~$S$ learns that the bit was received by~$R$.
We can abstract even further.  The reason that $S$ keeps sending
the bit to $R$ is that $S$ wants $R$ to know the value of the bit.
Thus, intuitively, $S$ should keep sending the bit until it knows that
$R$ knows its value.  Let $K_R({\it bit\/})$ be an abbreviation for
$K_R({\it bit\/} = 0) \lor K_R({\it bit\/} = 1)$, so $K_R({\it
bit\/})$ is true precisely if~$R$ knows the value of the bit.
The sender's behavior can be characterized by the
following knowledge-based program, $\PgbtK_S$:
\[\mbox{{\bf if} $K_SK_R(\bit)$ {\bf then} \noop~{\bf else}
{\sf sendbit}.}\]
Clearly when a message stating the value of the bit reaches
the receiver, $K_R(\bit)$ holds. But it also holds in other circumstances.
If, for example, the $K_SK_R(bit)$ holds initially, 
then there is no need to send anything. 
As above, it seems more efficient for the sender to stop sending
when he knows that the receiver 
will {\em eventually} know the value of the bit.
This suggests using the following program:
\[\mbox{{\bf if} $K_S(\doact(S,\noop)\rimp\Diamond K_R(\bit))$ 
{\bf then} \noop~{\bf else} {\sf sendbit}.}\]
However,
the same reasoning as in the case of $\PgbtR$ shows that this program 
is not implementable. 
And, again, using belief and counterfactuals, we can get a program 
$\PgbtB_S$ that does work, and uses fewer messages 
than~$\PgbtR_S$.
In fact, the following program does the job:
\[\mbox{{\bf if} $B_S (\doact(S,\noop) \RCond \Diamond B_R(\bit))$
{\bf then} \noop~{\bf else} {\sf sendbit},}\]
except that now we have to take $\B_R(\bit)$ to be an abbreviation for
$(\bit=0 \land B_R(\bit=0)) \lor (\bit = 1 \land B_R(\bit=1))$.
Note that 
$K_R(\bit)$, which was defined to be $K_R(\bit=0)) \lor K_R(\bit=1))$,
is logically equivalent to 
$(\bit=0 \land K_R(\bit=0)) \lor (\bit = 1 \land
K_R(\bit=1))$, since $K_R \phi \rimp \phi$ is valid for any formula
$\phi$.  But, in general, $B_R \phi \rimp \phi$ is not valid, so adding
the additional conjuncts in the case of belief makes what turns out to
be quite an important difference.  Intuitively, $B_R(\bit)$ says that~$R$ 
has correct beliefs about the value of the bit.

\commentout{
Similar problems were handled by Kislev and Moses
\cite{MosesKislev,Kislevthesis} in their work on
{\em knowledge-oriented programs}, also using an
approach that implicitly involves counterfactuals.
They define a high-level action called
{\sf Notify$(R,\varphi)$},
which is intended to ensure that $R$ knows~$\varphi$.
The implementation they had in mind says that when the notifier $S$
knows that {\em even if it
crashes (i.e., fails and
stops sending messages)\/} $R$ will
eventually know~$\varphi$, then $S$ can stop.
In the approach we consider here, $S$ is viewed, not as crashing, but as
running a program slightly different from the one it actually runs.
}%

The rest of this paper is organized as follows:  In the next section,
there is an informal review of the semantics of knowledge-based
programs.  
Section~\ref{counterfactuals} extends the knowledge-based framework 
by adding counterfactuals and beliefs.
We then formally analyze the programs $\PgbtR_S$ and $\PgbtB_S$, showing
that they have the appropropriate properties.
We conclude in
Section~\ref{discussion}.

\section{Giving semantics to knowledge-based programs}
Formal semantics for knowledge-based programs are provided by Fagin,
Halpern, Moses, and Vardi~\citeyear{FHMV,FHMV94}.
To keep the discussion in this paper at an informal
level, we simplify things somewhat here, and review
what we hope will be just enough of the details
so that the reader will be able
to follow the main points.
All the definitions in this section, except that of {\em de facto
implementation\/} at the end of the section, are taken
from~\cite{FHMV}.

Informally, we view a multi-agent system as consisting of a number
of interacting agents.  We assume that, at any given point in time,
each agent in the system is in some {\em local state}.
A {\em global state\/} is just a
tuple consisting of each agent's local state,
together with the state of the {\em environment}, where the
environment's state accounts for 
everything that is relevant to the system that
is not contained in the state of the processes.  The agents' local
states typically change over time, as a result of actions that they
perform.
A {\em run\/} is a function from time to global states.
Intuitively, a run is a complete description of what happens over time
in one possible execution of the system.  A {\em point\/} is a pair
$(r,m)$ consisting of a run $r$ and a time $m$.
If $r(m) = (\ells_e, \ells_1, \ldots, \ells_n)$, then we use $r_i(m)$ to
denote process $i$'s local state $\ell_i$ at the point $(r,m)$,
 $i = 1,\ldots, n$ 
and $r_e(m)$ to denote the environment's state $\ell_e$.  For
simplicity, time here is taken to range over the natural numbers 
rather than the reals
(so that time is viewed as discrete, rather than dense or continuous).   
{\em Round\/} $m$ in run $r$
occurs between time $m-1$ and $m$.   A {\em system\/} $\R$ is
a set of runs; intuitively, these runs describe all the
possible executions of the system.  For example, in a poker game, the
runs could describe all the possible deals and bidding sequences.

Of major interest in this paper are the systems that we can associate
with a program.  To do this, we must first associate a system with a
{\em joint protocol}.   A {\em protocol\/} is a function from local
states to 
nonempty sets of 
actions.
(We often consider {\em deterministic\/}  protocols, 
in which a local state is mapped to a singleton set of actions. Such protocols
can be viewed as functions from local states to actions.)
A joint protocol is just a set of protocols, one for
each process/agent.

We would like to be able to generate the system corresponding to a
given joint protocol~$P$.  To do this, we need to
describe the setting, or {\em context\/}, in which
$P$ is being 
executed.
Formally, a context $\gamma$ is a tuple $(P_e,\Gz,\tau,\Psi)$,
where $P_e$ is a protocol for the environment,
$\Gz$ is a set of initial global states,
$\tau$ is a {\em transition function}, and $\Psi$ is a set of
{\em admissible\/} runs.
The environment is viewed as running a protocol just like the agents;
its protocol is used to capture features of the setting 
such as ``all messages are delivered within 5 rounds'' or 
``messages may be lost''.
The transition function $\tau$
describes how the actions performed by the agents and the
environment change the global state by associating with each {\em joint
action\/} (a tuple consisting of an action for the environment and one
for each of the agents)
a {\em global state transformer}, that is, a mapping from global states 
to global states.
For the simple programs considered in this paper, the transition
function will be almost immediate from the description of the global
states.
The set~$\Psi$ of admissible runs is useful for capturing various 
fairness properties of the context. 
Typically, when no fairness constraints are imposed, $\Psi$ is
the set of all runs. 
(For a
discussion of the role of the set $\Psi$ of admissible runs
see~\cite{FHMV}.)
Since our focus in this paper is reasoning about actions and when 
they are performed,  we assume that all contexts are such that 
the environment's state at the point $(r,m)$ records the joint action
performed in the previous round (that is, between $(r,m-1)$ and $(r,m)$).
(Thus, we are essentially considering what are called {\em recording
contexts\/} in \cite{FHMV}.)

A run $r$ is consistent with a protocol $P$ if it could have been
generated when running protocol $P$.  Formally,
run $r$ is {\em consistent with joint
protocol $P$ in context $\gamma$\/} if
$r\in\Psi$ (so $r$ is admissible according to the context~$\gamma$), 
its  initial global state $r(0)$ is one of
the initial global states~$\Gz$ 
given in~$\gamma$, 
and for all~$m$, the transition
from global state $r(m)$ to $r(m+1)$ is the result of performing 
one of the joint actions specified by~$P$ and the environment protocol $P_e$
(given in~$\gamma$) in the global state $r(m)$.
That is, if $P = (P_1, \ldots, P_n)$, $P_e$ is the environment's
protocol in context $\gamma$, and
$r(m) = (\ell_e, \ell_1, \ldots, \ell_n)$, then
there must be a joint action $(\sfa_e, \sfa_1,\ldots, \sfa_n)$
such that $\sfa_e \in P_e(\ell_e)$, $\sfa_i \in P_i(\ell_i)$ for $i = 1,
\ldots, n$, and $r(m+1) = \tau(\sfa_e,\sfa_1,\ldots, \sfa_n)(r(m))$ (so
that $r(m+1)$ is the result of applying the joint action
$(\sfa_e,\sfa_1, \ldots, \sfa_n)$ to $r(m)$.
For future reference, we will say that a run~$r$ is {\em consistent
with $\gamma$} if $r$ is consistent with {\em some} joint protocol~$P$
in~$\gamma$. 
A system $\R$ {\em represents\/} a joint protocol $P$ in a context
$\gamma$
if it consists of all runs in $\Psi$ consistent with $P$ in $\gamma$.
We use $\Rrep(P,\gamma)$ to denote the system representing~$P$ in 
context~$\gamma$.

The basic logical language~$\L$ that we use is a standard
propositional temporal logic.
We start out with a set~$\Phi$ of primitive propositions~$p, q, \ldots$ 
(which are sometimes given more meaningful 
names such as $\recbit$ 
or $\recack$). Every primitive proposition is considered to be a
formula of~$\L$. We close off under the Boolean operators $\wedge$ 
(conjunction) and $\neg$ (negation). 
Thus, if $\phi$ and $\psi$ are formulas of~$\L$, then so are 
$\neg\phi$ and $\phi\wedge\psi$. 
The other Boolean operators are definable in terms
of these. E.g., implication $\phi\rimp\psi$ is defined as
$\neg(\neg\phi\wedge\psi)$. Finally, we close off under temporal
operators. For the purposes of this paper, it suffices to consider the
standard linear-time temporal operators $\next$ 
(``in the next (global) state')' and 
$\Diamond$ 
(``eventually''): 
If $\phi$ is a
formula, then so are  $\next\phi$ and $\Diamond\phi$.
The dual of $\Diamond$, which stands for ``forever,'' is denoted by 
$\Box$ and defined to be shorthand for $\neg\Diamond\neg$. 
This completes the definition of the language.

In order to assign meaning to the formulas of such a language~$\L$ in
a system~$\R$, we need an {\em interpretation} $\pi$, which determines
the truth of the primitive propositions at each of the 
global states 
of~$\R$. 
Thus, $\pi:\Phi\times \cG\to \{{\bf true}, {\bf false}\}$, 
where $\pi(p,g)={\bf true}$ exactly if the proposition~$p$ is true at
the global state~$g$. 
An {\em interpreted system} is a pair $\I=(\R,\pi)$ where $\R$ is a
system as before, and $\pi$ is an interpretation for~$\Phi$ in~$\R$. 
Formulas of~$\L$ are considered true or false at a point $(r,m)$ 
with respect to an interpreted system $\I=(\R,\pi)$ where $r\in\R$. 
Formally, 
\begin{itemize}
\item $(\I,r,m)\sat p$, for $p\in\Phi$, iff $\pi(p,r(m))={\bf true}$. 
\item $(\I,r,m)\sat \neg\phi$, iff $(\I,r,m)\not\sat \phi$.
\item $(\I,r,m)\sat \phi\wedge\psi$, iff both $(\I,r,m)\sat \phi$ and 
$(\I,r,m)\sat \psi$.
\item $(\I,r,m)\sat \next\phi$, iff $(\I,r,m+1)\sat \phi$.
\item $(\I,r,m)\sat \Diamond\phi$, iff $(\I,r,m')\sat \phi$ for some
$m'\ge m$.
\end{itemize}
By adding an interpretation~$\pi$ to the context~$\gamma$, we obtain an 
{\em interpreted context} $(\gamma,\pi)$.

We now describe a simple programming language, introduced in~\cite{FHMV}, 
which is still rich enough to describe protocols, and whose syntax
emphasizes the fact that an agent performs actions based on the result
of a test that is applied to her local state.
A ({\em standard}\index{standard program}) {\em program}
for agent~$i$ is a statement of the form:
\begin{tabbing}
\ \ \ \ {\bf case} \= {\bf of}  \\
\> {\bf if} $t_1 $ {\bf do} $\sfa_1$  \\
\> {\bf if} $t_2 $ {\bf do} $\sfa_2$  \\
\> $\cdots$  \\
\ \ \ \ {\bf end case}
\end{tabbing}
where the $t_j$'s are {\em standard tests\/}\index{standard test}
for agent~$i$
and the $\sfa_j$'s are actions of agent~$i$ (i.e., $\sfa_j \in \ACT_i$).
(We later modify these programs to obtain {\em knowledge-based\/} and
{\em belief-based} programs; the distinction will come from the kinds of tests 
allowed.   
We omit the {\bf case} 
statement if there is only one clause.)
A standard test for agent~$i$ is simply a propositional formula
over a set $\Phi_i$ of primitive propositions.
Intuitively, if $L_i$ represents the local states of agent~$i$ in~$\cG$, 
then once we know how to evaluate the tests in the program at the local
states in~$L_i$, we can convert this program to a protocol over~$L_i$:~at
a local state~$\ells$, agent~$i$ nondeterministically chooses
one of the (possibly infinitely many) clauses in the {\bf case\/}
statement whose test is true at~$\ells$, and executes the corresponding
action.

We want to use an interpretation~$\pi$ to tell us how to
evaluate the tests.
However, not just any interpretation will do.
We intend the tests in a program for agent~$i$ to be
{\em local}\oldindex{local test}, that is,
to depend only on agent~$i$'s local state.
It would be inappropriate
for agent~$i$'s action to depend on the truth value of a test that~$i$
could not determine from her local state.
An interpretation~$\pi$ on the global states in~$\G$ is
{\em compatible\/} with a program $\Pg_i$ for agent~$i$ if
every proposition that appears in $\Pg_i$ is local to~$i$; 
that is, if $q$ appears in~$\Pg_i$,
the states $s$ and~$s'$ are in~$\G$,
and $s \sim_i s'$, then $\pi(s)(q) = \pi(s')(q)$.
If $\phi$ is a
propositional formula all of whose primitive propositions are local to
agent~$i$, and~$\ells$ is a local state of agent~$i$, then we write
$(\pi,\ells) \sat \phi$ 
if $\phi$ is satisfied by the truth assignment
$\pi(s)$, where $s = (s_e, s_1, \ldots, s_n)$ is a global state such
that $s_i = \ells$.
 Because all the primitive propositions in~$\phi$ are
local to~$i$, it does not matter which global state~$s$ we choose, as
long as $i$'s local state in~$s$ is~$\ells$.
Given a
program
$\Pg_i$ for agent~$i$ and an
interpretation~$\pi$ compatible with~$\Pg_i$, we
define a protocol that we denote $\Pg_i^\pi$
by setting:
\[ \Pg_i^\pi(\ells) = \left\{
      \begin{array}{lll}
       \{\sfa_j \stc (\pi, \ells) \sat t_j \}
       & \mbox{if } \{j \stc (\pi, \ells) \sat t_j \} \ne \emptyset \\[0.5ex]
       \{ \noop\}
      & \mbox{if } \{j \stc (\pi, \ells) \sat t_j \} = \emptyset .
\end{array}
\right.
\]
Intuitively, $\Pg_i^\pi$ selects all actions from the clauses that
satisfy the test, and selects the null action~$\noop$ if no test
is satisfied.
In general, we get a nondeterministic protocol, since more than
one test may be satisfied at a given state.

Many of the definitions that we gave for protocols have natural
analogues for programs.
We define a {\em joint\/} program\index{joint program}
to be a tuple $\Pg=(\Pg_1,\ldots,\Pg_n)$, where $\Pg_i$ is
a program for agent~$i$.
An interpretation $\pi$ is
{\em compatible\/} with~$\Pg$ if~$\pi$ is compatible with each of the
$\Pg_i$'s. 
{F}rom~$\Pg$ and~$\pi$ we get
a joint protocol $\Pg^\pi = (\Pg_1^\pi, \ldots, \Pg_n^\pi)$. %
\glossary{\glospgpi}
We say that
an interpreted system $\I = (\R,\pi)$ {\em represents\/} 
a joint program $\Pg$ in the interpreted
context $(\gamma,\pi)$ exactly if~$\pi$ 
is compatible with~$\Pg$ and~$\I$ 
represents the corresponding protocol $\Pg^\pi$.
We denote the interpreted system representing $\Pg$ in $(\gamma,\pi)$ by
$\Irep(\Pg,\gamma,\pi)$\glossary{\glosipgg}.
Of course, this definition only makes sense if $\pi$ is compatible 
with~$\Pg$.  {F}rom now on we always assume that this is the case. %
\index{interpreted system!represents standard program|)}

The syntactic form of our standard programs is in many ways more
restricted than that of programs in common programming languages such
as C or FORTRAN. In such languages, one typically sees
constructs such as ${\bf for}$, ${\bf while}$,
or \mbox{{\bf if\ldots then\ldots else\ldots}},
which do not have syntactic analogues in our formalism.
As discussed in~\cite{FHMV}, it is possible to encode a program
counter in tests and actions of standard programs. By doing so, 
it is possible to simulate these constructs. Hence, there is
essentially no loss of generality in our definition of standard programs.
Since 
each test in a standard program $\Pg$ run by process $i$
can be evaluated in each local state, we can derive a
protocol from $\Pg$ in an obvious way:~to find out what process $i$
does in a local state $\ell$, we evaluate the
tests in the program in $\ell$ and perform the appropriate action.
A run is {\em consistent with $\Pg$ in interpreted 
context $(\gamma,\pi)$\/} if it is consistent with the protocol
derived from $\Pg$.  Similarly, a system {\em represents $\Pg$ in
interpreted context $(\gamma,\pi)$\/} if it represents the protocol 
derived from $\Pg$ in $(\gamma,\pi)$.
\xam\label{xam0}
Consider the (joint) program $\Pgbt = (\Pgbt_S,\Pgbt_R)$,
where $\Pgbt_S$ is as defined in the introduction, and $\Pgbt_R$ is the
program 
$$\mbox{{\bf if} $\recbit$ {\bf then} ${\sf sendack}$ {\bf else}
$\noop$}.$$
Thus, in $\Pgbt_R$, 
the receiver sends an acknowledgement if it has received 
the bit, 
and otherwise does nothing.  
This program, like all the programs considered in this paper,
is applied 
repeatedly, 
so it effectively runs forever.
Assume 
that
$S$'s local state includes the time, its input bit, and whether
 or not~$S$ has received an acknowledgment from $R$; the state
thus has the form
$(m,i,x)$, where $m$ is a natural number (the time), $i \in \{0,1\}$ is
the input bit, and $x \in \{\lambda,\ack\}$.  Similarly,
$R$'s local state has the form $(m,x)$, where $m$ is the time and $x$ is
either
$\lambda$,
0, or 1, depending
on whether or not it has received the bit from $S$ and what the bit is.
As in all recording contexts, the environment state keeps track of the
actions performed by the agents.  Since the environment state plays no
role here, we omit it from the description of the global state,
and just identify the global state with the pair consisting of $S$ and
$R$'s local state.
Suppose that, in context $\gamma$,
the environment protocol
nondeterministically decides whether or not a
message sent by $S$ and/or $R$ is delivered, the initial global states
are $((0,0,\lambda),(0,\lambda))$ and 
$((0,1,\lambda),(0,\lambda))$, the
transition function is such
that the joint actions have the obvious effect on the global state, and
all runs are admissible.  
Then a run consistent with $\Pgbt$ in $(\gamma,\pi)$ in which $S$'s
bit is 0, $R$ receives the bit in the second round, and $S$ receives an
acknowledgment from $R$ in the third round has the
following sequence of global states: $((0,0,\lambda),(0,\lambda)),
((1,0,\lambda),(1,\lambda)),
((2,0,\lambda),(2,0)), ((3,0,\ack),(3,0)),
((4,0,\ack),(4,0)), \ldots$.
\exam

Now we consider knowledge-based programs.
We start by extending our logical language by adding 
a modal operator $K_i$ for every agent $i=1,\ldots,n$.
Thus, whenever $\phi$ is a formula, so is $K_i\phi$. 
Let $\Lk$ be the resulting language.
According to the standard definition of knowledge in systems \cite{FHMV},
an agent $i$ knows a fact
$\phi$ at a given point $(r,m)$ in interpreted system $\I=(\R,\pi)$ 
if $\phi$ is true at all points in $\R$ 
where~$i$ has the same local state as it does at $(r,m)$.
We now have 
\begin{itemize}
\item $(\I,r,m) \sat K_i \phi$ if $(\I,r',m') \sat \phi$ for all points
$(r',m')$ such that $r_i(m) = r'_i(m')$.
\end{itemize}
Thus, $i$ knows $\phi$ at the point $(r,m)$ if $\phi$
holds at all points consistent with $i$'s information at $(r,m)$.

A {\em knowledge-based program} has the same structure as a 
standard program except that all tests in the program 
text~$\Pg_i$  for agent~$i$ are formulas of the form $K_i\psi$.%
\footnote{All standard programs can be viewed as knowledge-based
programs.  Since all the tests in a standard program for agent $i$
must be local to $i$, every test $\phi$ in a standard program for agent
$i$ is equivalent to $K_i \phi$.}
As for standard programs, we can 
define when a protocol 
implements a knowledge-based program, except this time 
it is with respect to an  interpreted context. 
The situation in this case is, however, somewhat more complicated.
In a given context, a process can determine the truth of a
standard test such as ``$x=0$'' by simply 
checking
its local state.
However, the truth of the tests for knowledge that appear in
knowledge-based programs  cannot in general
be determined simply by looking at the local state in isolation.  
We need to look at the whole system.  
As a consequence,
given a run, we cannot in general determine if it is consistent with a
knowledge-based program in a given interpreted context.  
This is because we cannot tell how the tests for knowledge turn out without
being given the other possible runs of the system; what a process knows
at one point will depend in general on what other points are possible.
This stands in sharp contrast to the situation for standard programs.

This means it no longer makes sense to talk about a run being consistent
with a knowledge-based program in a given context.
However, notice that, given an interpreted system $\I=(\R,\pi)$, we
can derive a protocol from a knowledge-based program $\Pgkb$ for
process $i$ by
evaluating the knowledge tests in $\Pgkb$ with respect to~$\I$.
That is, a test such as $K_i \phi$ holds in a local state $\ell$ if 
$\phi$ holds at all points $(r,m)$ in $\I$ such that 
$r_i(m) = \ell$.%
\footnote{Note that if
there is no point $(r,m)$ in $\I$ such that $r_i(m) = \ell$, 
then $K_i \phi$ 
vacuously holds at~$\ell$, for all formulas~$\phi$.} 
In general, different protocols can be derived from a
given knowledge-based program, depending on what system
we use to evaluate the tests.
Let $\Pgkb^\I$
denote the protocol derived from $\Pgkb$ 
by using $\I$ to evaluate the tests for knowledge.
An interpreted system $\I$ {\em represents\/}
the knowledge-based program
$\Pgkb$ in interpreted context $(\gamma,\pi)$ if $\I$ represents 
the protocol $\Pgkb^\I$.
That is, $\I$ represents $\Pgkb$ if 
$\I = \Irep(\Pgkb^\I,\gamma,\pi)$.  Thus, a system represents $\Pgkb$ if it
satisfies a certain fixed-point equation.
A protocol $P$ {\em implements\/} $\Pgkb$ in interpreted context
$(\gamma,\pi)$ 
if $P = \Pgkb^{\Irep(P,\gamma,\pi)}$. 

This definition
is somewhat subtle, and determining
the protocol(s) implementing a given
knowledge-based program may be nontrivial.
Indeed, as shown by
Fagin, Halpern, Moses, and Vardi~\citeyear{FHMV,FHMV94},
in general, there may be no
protocols implementing a knowledge-based
program~$\Pgkb$ in a given context, there may be only one, or there may
be more than one, since
the fixed-point equation may have no solutions, one solution, or many
solutions.   In particular, it is not hard to show that
there is no
(joint)
protocol implementing
a (joint) program where $S$ uses $\Pgbt_S^*$ or $\Pgbt_S^{\rimp}$, as
described in the introduction.
For the purposes of this paper, it is useful to have a notion slightly
weaker than that of implementation.
Two joint protocols $P = (P_1, \ldots, P_n)$ and $P' = (P'_1, \ldots, P'_n)$
are {\em equivalent in context $\gamma$}, denoted 
$P \approx_\gamma P'$, if 
(a) $\Rrep(P,\gamma) = \Rrep(P',\gamma)$ and 
(b) $P_i(\ell) = P'_i(\ell)$ for every local state $\ell=r_i(m)$ with
$r \in \Rrep(P,\gamma)$.
Thus, two protocols that are equivalent in~$\gamma$ may disagree on the
actions performed in some local states, 
provided that those local states never arise in the actual runs of these 
protocols in~$\gamma$.
We say $P$ {\em de facto implements\/} a knowledge-based program $\Pgkb$ in
interpreted context $(\gamma,\pi)$ if $P \approx_\gamma
\Pgkb^{\Irep(P,\gamma,\pi)}$.  
Arguably, de facto implementation suffices for most purposes, since all we
care about are the runs generated by the protocol.  We do not care about
the behavior of the protocol on local states that never arise.

It is almost immediate from the definition that if $P$ implements
$\Pgkb$, then $P$ de facto implements $\Pgkb$.  The converse may not be
true, since we may have 
$P \approx_\gamma \Pgkb^{\Irep(P,\gamma,\pi)}$ without
having $P = \Pgkb^{\Irep(P,\gamma,\pi)}$. On the other hand, as the
following lemma shows, if $P$ de facto implements $\Pgkb$, then a
protocol closely related to $P$ implements $\Pgkb$.

\lem\label{weakimp} If $P$ de facto implements $\Pgkb$ in $(\gamma,\pi)$ then
$\Pgkb^{\Irep(P,\gamma,\pi)}$ implements $\Pgkb$ in $(\gamma,\pi)$.
\elem

\prf Suppose that $P$ de facto implements $\Pgkb$ in $(\gamma,\pi)$.  
Let $P' = \Pgkb^{\Irep(P,\gamma,\pi)}$.  
By definition, $P' \approx_\gamma P$.  
Thus, $\Irep(P',\gamma,\pi) = \Irep(P,\gamma,\pi)$, so $P' =
\Pgkb^{\Irep(P',\gamma,\pi)}$.
It follows that~$P'$ implements~$\Pgkb$.  
\eprf

\section{Counterfactuals and Belief}
\label{counterfactuals}
In this section, we show how counterfactuals and belief can be added to
the knowledge-based framework, and use them to do a formal analysis of
the programs $\PgbtR_S$ and $\PgbtB_S$ from the introduction.

\subsection{Counterfactuals}

The semantics we use for counterfactuals is
based on the standard semantics used in the philosophy
literature~\cite{Lewis73,Stalnaker68}.  As with other modal logics, this
semantics starts 
with a set $W$ of possible worlds.  For every
possible world~$w \in W$ there is a (partial) order $\lt_w$ defined 
on $W$.  Intuitively, $w_1 \lt_w w_2$ if $w_1$ is ``closer'' or 
``more similar'' to world~$w$ than $w_2$ is.
This partial order is assumed to satisfy certain constraints, such as 
the condition that $w \lt_w w'$ for all $w' \ne w$: world~$w$ is closer 
to~$w$ than any other world is.  
A counterfactual statement of the form
$\phi \RCond \psi$ is then taken to be true at a world~$w$ if, in all
the worlds closest to~$w$ 
among the worlds 
where $\phi$ is true, $\psi$ is also true.

In our 
setting,
we  obtain a notion of closeness by 
associating with every point $(r,m)$ of a system~$\cI$ a partial 
order on the points of~$\cI$.%
\footnote{In a more general treatment, we could 
associate a different partial order with 
every agent at every point; this 
is not
necessary for 
the examples we consider
in this paper.}
An {\em order assignment\/} for a system~$\cI=(\R,\pi)$ 
is a function~$\lta$ that associates with every point~$(r,m)$ of~$\cI$ 
a 
partial order relation~$\lta_{(r,m)}$ over the points of~$\cI$. 
The partial orders must satisfy the constraint that $(r,m)$ is a 
minimal element of~$\lta_{(r,m)}$, so that there is no 
run $r'\in\R$ and time $m'\ge 0$ satisfying $(r',m')\,\lta_{(r,m)}(r,m)$. 
A {\em counterfactual system} is a pair of the form 
$\cJ=(\cI,\lta)$, 
where $\cI$ is an interpreted system as before, while~$\lta$ is an order 
assignment for the points in $\cI$.
Given a counterfactual system $\cJ = (\cI,\lta)$, a point $(r,m)$ in
$\cI$, and a set~$A$ of points of~$\cI$, define 
\[
\begin{array}{ll}
\closest(A,(r,m),\cJ)~~ =\cr
 \qquad~~\{ (r',m') \in A:\,
\mbox{there is no $(r'',m'') \in A$ such that 
$(r'',m'') \, \lta_{(r,m)}\,(r',m')\}$.}
\end{array}
\]
Thus, $\closest(A,(r,m),\cJ)$ consists of the 
closest points to $(r,m)$ among the points in~$A$ 
(according to the order assignment $\lta$). 

To allow for counterfactual statements, we extend our logical 
language~$\L$ with a binary operator $\RCond$ on formulas, so that whenever
$\phi$ and~$\psi$ are formulae, so is $\phi\RCond\psi$. We read 
$\phi\RCond\psi$ as ``{\sl if~$\phi$ were the case, then}~$\psi$,''
and denote the resulting language by $\Lc$. 

Let~$\intension{\phi}  = \{(r,m): (\cJ,r,m) \sat \phi\}$; that is,
$\intension{\phi}$ consists of all points in $\cJ$ satisfying $\phi$.
We can now define the semantics of counterfactuals as follows: 
\[
(\cJ,r,m)\sat \phi \RCond \psi~~{\rm if~~} (\cJ,r',m')\sat\psi ~~
\mbox{for all $(r',m')\in\closest(\intension{\phi},(r,m),\cJ)$}.
\] 
This definition captures the intuition for counterfactuals stated
earlier: $\phi \RCond \psi$ is true at a point $(r,m)$ if $\psi$ is true
at the points closest to $(r,m)$ where $\phi$ is true.

All earlier analyses of  (epistemic) properties of a protocol $P$ in a 
context $\gamma$ used the interpreted system $\Irep(P,\gamma,\pi)$,
consisting of all the runs consistent 
with~$P$ in context~$\gamma$.
However, counterfactual reasoning involves events that occur on runs that
are not consistent with $P$.  To support such reasoning we need to 
consider runs not in $\Irep(P,\gamma,\pi)$.  
The runs that must be added
can, in  
general, depend on the type of counterfactual statements allowed in 
the logical language. Thus, for example, if we allow formulas of 
the form $\doact(i,\sfa)\RCond\psi$ for process~$i$ and action~$\sfa$, then 
we must allow, at every point of the system, a possible future in which~$i$'s
next action is~$\sfa$.%
\footnote{Recall from the introduction that our programs use the formula 
$\doact(i,{\sf a})$ to state that agent~$i$
is about to perform  action~${\sf a}$.
Thus,  $\doact(i,{\sf
a})\RCond\phi$ says ``if agent~$i$ were to perform {\sf a} then $\phi$ 
would be the case.'' We assume that all interpretations we consider give
this formula the appropriate meaning.
If the protocol $P$ being used is encoded in the global state (for example,
if it is part of the environment state),
then we can take  $\doact(i,{\sf a})$ to be a primitive proposition.  
Otherwise, we cannot, since its truth cannot be determined 
from the global state.  However, we can always take $\doact(i,{\sf a})$
to be an abbreviation for $\next\latest(i,{\sf a})$, where the
interpretation $\pi$ ensures that
$\latest(i,{\sf a})$ is true at a point $(r,m)$ if~$i$ 
performed ${\sf a}$  in round~$m$ of~$r$.   Since we assume the the last
joint action performed is included in the environment state, the truth
of $\latest(i,{\sf a})$ is determined by the global state.}

An even richer set of runs is needed if we allow 
the language to specify a sequence of actions performed by a given process, 
or if counterfactual conditionals~$\RCond$ can be nested.
To handle a broad class of applications, including ones involving 
formulas with temporal operators and arbitrary nesting 
of conditional statements involving $\doact(i,\sfa)$,
we do reasoning with respect to the
system~$\Ip(\gamma,\pi)=(\Rp(\gamma),\pi)$ 
consisting of {\em all} 
runs compatible~$\gamma$, that is, all runs consistent with some
protocol $P'$ in context $\gamma$.
In this way all possible behaviors, within the constraints induced
by~$\gamma$, can be reasoned about. 
There is a potential problem with using
system~$\Ip(\gamma,\pi)=(\Rp(\gamma),\pi)$ for reasoning about $P$:~all
reference to $P$ has been lost.  We return to this issue in the next
section, when we discuss belief.  For now we show how to use
$\Ip(\gamma,\pi)$ as a basis for doing counterfactual reasoning.

As we have already discussed, the main issue in using $\Ip(\gamma,\pi)$
to reason about $P$ is that of defining an appropriate order assignment.
We are interested in order assignments that depend on the protocol in a
uniform way.
An {\em order generator~$\ogen$} for a context~$\gamma$ 
is a function that associates with every protocol~$P$ 
an order assignment $\lta^P=\ogen(P)$ on the points of $\Rp(\gamma)$. 
A {\em counterfactual context} is a tuple 
$\zeta=(\gamma,\pi,\ogen)$, where~$\ogen$ is an order generator
for~$\gamma$. 
In what follows we denote by 
$\Jc(P,\zeta)$ the counterfactual system $(\Ip(\gamma,\pi),\ogen(P))$,
where $\zeta=(\gamma,\pi,\ogen)$; we omit~$\zeta$ when it is clear
from context.
We are interested in order generators $\ogen$ such that $\ogen(P)$ says
something about deviations from $P$.
For the technical results we prove in the rest of the paper, 
we focus on order generators that prefer runs in which the agents do 
not deviate from their protocol. 
Given an agent $i$, action $\sfa$, protocol $P$, context $\gamma$, and
point $(r,m)$ in $\Rp(\gamma)$, define 
$\close(i,\sfa,P,\gamma,(r,m)) =
\{(r',m):$ (a) $r' \in \Rp(\gamma)$,
(b) $r'(m') = r(m')$ for all $m' \le m$, 
(c) if agent $i$ performs ${\sf a}$ in round~$m+1$ of~$r$, then $r' = r$,
(d) if agent~$i$ does not perform perform ${\sf a}$ in round~$m+1$
of~$r$, then agent $i$ 
performs ${\sf a}$ in round~$m+1$ of~$r'$ and 
follows~$P$ in all later rounds, 
(e) all agents other than $i$ follow $P$ from round $m+1$ on in $r'\}$.
That is, $\close(i,\sfa,P,\gamma,(r,m))$ is 
the set of points $(r',m)$ where run $r'=r$ if $i$ performs $\sfa$ in
round $m+1$ of $r$; otherwise, $r'$ 
is identical to $r$ up to time $m$ and all the agents act according to
joint protocol $P$ at all 
later
times, except that at the point $(r',m)$,
agent $i$ performs action~{\sf a}. 
An order generator~$\ogen$ for~$\gamma$ 
{\em respects protocols\/} if, for every protocol~$P$,
point $(r,m)$ of~$\Rrep(P,\gamma)$, action~{\sf a}, and agent~$i$, 
$\closest(\intension{\doact(i,{\sf a})},(r,m),\Jc(P))$ 
is a nonempty subset of
$\close(i,\sfa,P,\gamma,(r,m))$
that includes $(r,m)$.
Of course, the most obvious order generator that respects protocols just
sets $\closest(\intension{\doact(i,{\sf a})},(r,m),\Jp(P)) =
\close(i,\sfa,P,\gamma,(r,m))$.
Since our results hold for arbitrary order
generators that respect protocols, we have allowed the extra 
flexibility of allowing $\closest(\intension{\doact(i,{\sf
a})},(r,m),\Jp(P))$ to be a strict subset of
$\close(i,\sfa,P,\gamma,(r,m))$. 

A number of points are worth noting about this definition:
\begin{itemize}
\item 
If the environment's protocol~$P_e$ and the agents' individual protocols
in~$P$ are all deterministic, then 
$\close(i, \sfa, P, \gamma, (r,m))$ 
is a singleton, since there is a unique run where the agents act 
according to joint protocol $P$ at all times except that agent~$i$
performs action {\sf a} at time $m$.  Thus,
$\closest(\intension{\doact(i,{\sf a})},(r,m),\Jc(P))$ must be the
singleton $\close(i, \sfa, P,\gamma, (r,m))$ in this case.
However, in many cases, it is
best
to view the environment as following a nondeterministic protocol
(for example, nondeterministically deciding at which round a message
will be delivered); in this case, there may be several 
points in $\cI$ closest to
$(r,m)$.  Stalnaker \citeyear{Stalnaker68} required there to be a unique
closest world; Lewis \citeyear{Lewis73} did not.  There was later
discussion of how reasonable this requirement was (see, for example,
\cite{Stalnaker80}).  
Thinking in terms of systems may help inform this debate.
\item 
If process $i$ does not perform action ${\sf a}$ at the point
$(r,m)$, then there may be points in
$\closest(\intension{\doact(i,{\sf a})},(r,m),\Jc(P))$
that are not in $\Rrep(P,\gamma)$,
even if $r \in \Rrep(P,\gamma)$.
These 
points are ``counter to fact''.
\item 
According to our definition, the notion of ``closest'' depends on
the protocol that generates the system.
For example, consider a context $\gamma'$ that is just like the 
context~$\gamma$ from Example~\ref{xam0}, except that $S$ keeps track in its
local state, not only of the time, but also of the number of messages
it has sent. 
Suppose that the protocol $P_S$ for $S$ is determined by the program
$$\mbox{{\bf if} \clock=0 {\bf then} $\sf sendbit$ {\bf else}
$\noop$,}$$
while $P_S'$ is the protocol 
determined by the program
$$\mbox{{\bf if} {\em \#messages}=0 {\bf then} $\sf sendbit$ {\bf
else} $\noop$}.$$
Let $P = (P_S,\NOOP_R)$ and $P' = (P_S',\NOOP_R)$, where $P_R$ is the protocol
where $R$ does nothing (performs the action~$\noop$) in all states.
Clearly $\Rrep(P,\gamma') = \Rrep(P',\gamma')$: whether it is following
$P_S$ or $P_S'$, the sender $S$ sends a message only in the first round
of each run.  
It follows that these two protocols specify exactly the same 
behavior in this context. 
While these protocols coincide when no deviations take place, 
they may differ if deviations are possible.
For example, imagine a situation where, for whatever reason, $S$ did
nothing in the first round. In that case, at the end of the first round, 
the clock has advanced from~0, while the count of the number of
messages that~$S$ has sent is still~0.  
$P$ and~$P'$ would then produce different behavior in the
second round. This difference is captured by our definitions. 
If~$\ogen$ respects protocols, then 
$\closest(\intension{\doact(S,\noop)},(r,0),\Jc(P))\ne 
\closest(\intension{\doact(S,\noop)},(r,0),\Jc(P'))$.
No messages are sent by $S$ in runs appearing in points in 
$\closest(\intension{\doact(S,\noop)},(r,0),\Jc(P))$, while 
exactly one message is sent by $S$ in each run appearing in points in
$\closest(\intension{\doact(S,\noop)},(r,0),\Jc(P'))$.

This dependence on the protocol is a deliberate feature of our definition;
by using order generators, the order assignment we consider is a
function of the protocol being used.
While the protocols~$P$ and~$P'$ specify the same behavior
in~$\gamma$, they specify different behavior 
in ``counterfactual'' runs, 
where something happens that somehow causes behavior inconsistent with
the protocol.  The subtle difference between the two protocols is 
captured by our definitions.
\end{itemize}

\subsection{Belief}
As we have just seen, in order to allow for counterfactual reasoning about a 
protocol~$P$ in a context~$\gamma$, our model needs to 
represent ``counterfactual''  runs that do not appear in~$\Rrep(P,\gamma)$. 
Using the counterfactual system~$\Jc(P)$, which includes all runs
of~$\Rp(\gamma)$,  
provides considerable flexibility 
and generality in counterfactual reasoning. 
However, doing this has a rather drastic impact on the processes'
knowledge of the protocol being used.
Agents have considerable knowledge of the properties of protocol~$P$
in the interpreted system $\Irep(P,\gamma)$, since it
contains only the runs of~$\Rrep(P,\gamma)$. For example, if  
agent~1's first action in~$P$ is always~${\sf b}$, 
then all agents are guaranteed to know 
this fact (provided that it is expressible in the language, of
course); indeed, this fact will be {\em common knowledge}, which means
agent knows it, for any depth of nesting of these knowledge statements
(\cf\cite{FHMV,HM90}).
If we evaluate knowledge with respect to~$\Rp(\gamma)$, 
then the agents have lost
the knowledge that they are running protocol~$P$.
We deal with this by adding extra information to the models that allows
us to capture the agents' beliefs.  Although the agents will not {\em
know\/} they are running protocol $P$, they will {\em believe\/} that
they are.
A {\em ranking function\/} for a system~$\R$ is a function 
$\rank:\R\to\Natp$, associating with every run of~$\R$
a {\em rank}~$\rank(r)$, which is either a natural number or~$\infty$,
such that $\min_{r \in \R} \rank(r) = 0$.%
\footnote{The similarity in notation with the $\kappa$-rankings of
\cite{Goldszmidt92}, which are based on Spohn's {\em ordinal conditional
functions\/} \citeyear{spohn:88}, is completely intentional.  Indeed,
everything we are saying here can be recast in Spohn's framework.}
Intuitively, the rank of a run defines the likelihood of the run. Runs
of rank~0 are most likely; runs of rank~1 are somewhat less likely, those
of rank~2 are even more unlikely, and so on. 
Very roughly speaking, if~$\epsilon>0$ is small, we can think of 
the runs of rank~$k$ as having probability $O(\epsilon^k)$.
For our purposes, the key feature of rankings is that they can be used
to define a notion of belief
(cf.~\cite{FrH1Full}).
Intuitively, of all the points considered possible 
by a given agent at a point $(r,m)$, the ones believed to have
occurred are the ones appearing in runs of minimal rank. 
More formally, for a point $(r,m)$ define 
\[\mini^\rank(r,m)~=~\min\{\rank(r')\,|\, r'\in\R(\gamma)~{\rm and}
~r'_i(m')=r_i(m) \mbox{~for some $m'\ge 0$}\}.\]
Thus, $\mini^\rank(r,m)$ is the minimal $\rank$-rank 
of runs in which~$r_i(m)$
appears as a local state for agent~$i$. 
An {\em extended system} 
is a triple of the form 
$\cJ=(\cI,\lta,\rank)$, where $(\cI,\lta)$ is a counterfactual system, 
and $\rank$ is a ranking function for the runs of~$\I$. 
In extended systems we can define a notion of belief.
The logical language that results from closing~$\Lc$ (resp.~$\L$) under belief
operators~$B_i$, for $i=1,\ldots,n$, is denoted~$\Lbc$ (resp.~$\Lb$).
The truth of $B_i\phi$ is defined as follows:
\[\begin{array}{ll}
(\I,\lta,\rank,r,m)\sat B_i\phi~ \mbox{ iff } 
&(\I,\lta,\rank,r',m')\sat \phi
~\mbox{for all $(r',m')$ such that}\\
&\mbox{$\rank(r')=\mini^\rank(r,m)$ and 
$r'_i(m')=r_i(m)$.}
\end{array}\]
What distinguishes knowledge from belief is that knowledge satisfies the
{\em knowledge axiom}: $K_i \phi \rimp \phi$ is valid.   While $B_i \phi
\rimp \phi$ is not valid, it is true in runs of rank~0.

\lem\label{knowledge-axiom}
Suppose that $\cJ=((\R,\pi),\lta,\rank)$ is an extended system, $r \in
\R$, and $\rank(r) = 0$.
Then for every formula~$\phi$ and all times~$m$, we have
$(\cJ,r,m)\sat B_i\phi\rimp\phi$.
\elem
\prf
Assume that $\rank(r) = 0$.
Thus, $\mini^\rank(r,m)=0$ 
for all times $m\ge 0$. 
It now immediately follows from the definitions that if 
$(\cJ,r,m)\sat B_i\phi$, then  $(\cJ,r,m)\sat \phi$.
\eprf

By analogy with order generators, we now want a uniform way of
associating with each protocol $P$ a ranking function.  Intuitively, we
want to do this in a way that lets us recover $P$.
We say that a ranking function~$\rank$ is {\em $P$-compatible} 
(for~$\gamma$) if $\rank(r)=0$ if and only if $r\in\Rrep(P,\gamma)$.
A {\em ranking generator} 
for a context~$\gamma$ is a
function~$\rgen$ ascribing to every protocol~$P$ a ranking
$\rgen(P)$ on the runs of $\Rp(\gamma)$.  
A ranking generator~$\rgen$ is 
{\em deviation compatible\/} 
if $\rgen(P)$ is $P$-compatible for every protocol~$P$. 
An obvious example of a deviation-compatible ranking generator is the 
{\em characteristic} ranking generator $\rgen_\xi$
that, for a given protocol~$P$, yields a ranking that 
assigns rank $0$ to every run in~$\Rrep(P,\gamma)$ and 
rank~$1$ to all other runs. 
This captures the assumption that runs of~$P$ are likely 
and all other runs are unlikely, without attempting to distinguish 
among them. 
Another deviation-compatible ranking generator is $\rgend$, 
where the ranking $\rgend(P)$ assigns to a run~$r$ the total number
of times that agents deviate from~$P$ in~$r$. 
Obviously, $\rgend(P)$ assigns~$r$ the rank~0 exactly if 
$r\in\Rrep(P,\gamma)$, as desired. 
Intuitively, $\rgend$ captures the assumption that not only are
deviations unlikely, but they are independent. 
It is clearly possible to construct other $P$-compatible rankings that
embody other assumptions.
For example, deviation can taken to be an indication of faulty
behavior.  Runs of rank $k$ can be those where exactly $k$ processes are
faulty. 

Our interest in deviation-compatible ranking generators
is motivated by the observation that the notion of belief
that they give rise to in~$\Ip(\gamma,\pi)$ 
generalizes 
the notion of 
knowledge with respcet to $\Irep(P,\gamma,\pi)$.  
To make this precise, define $\phi^B$ to be the formula 
that 
is obtained
by replacing all $K_i$ operators in~$\phi$ by $B_i$. 
(Notice that if $\phi\in\Lk$ then $\phi^B\in\Lb$.)
In addition, since ranking generators now play a role in determining 
beliefs, we define an {\em interpreted belief context} to be a triple 
of the form $(\gamma,\pi,\rgen)$. 

\thm\label{genthm}
Let $\rgen$ be a deviation-compatible ranking generator for~$\gamma$.
For every formula $\phi\in\Lk$ and for all points 
$(r,m)$ of $\R=\Rrep(P,\gamma)$ 
and every ordering~$\lta$
we have 
\[ (\Irep(P,\gamma,\pi),r,m)\sat\phi~~\mbox{iff}~~ 
(\Ip(\gamma,\pi),\lta,\rgen(P),r,m)\sat\phi^B.\]
\ethm

\prf 
We proceed by induction on the structure of $\phi$.  For primitive
propositions, the result is immediate by definition, and the argument is
trivial if $\phi$ is a conjunction or a negation.  Thus, assume that
$\phi$ is of the form $K_i\psi$.  
Let $\rank=\rgen(P)$.
Then 
$(\Irep(P,\gamma,\pi),r,m)\sat K_i\psi$ iff
$(\Irep(P,\gamma,\pi),r',m')\sat\psi$ for all $(r',m')$ such that $r' \in
\Rrep(P,\gamma)$ and $r'_i(m') = r_i(m)$.  But $r' \in \Rrep(P,\gamma)$
iff $\kappa(r') = 0$.  Thus, $(\Irep(P,\gamma,\pi),r,m)\sat K_i\psi$ iff
$(\Ip(\gamma,\pi),r',m')\sat\psi^B$ for all $(r',m')$ such that $\rank(r')
= 0$ and $r'_i(m') = r_i(m)$.  
Note that
$\mini^\rank(r,m)~= 0$ 
(because
$\rank(r) = 0$).  Thus, it easily follows that $(\Irep(P,\gamma,\pi),r,m)\sat
K_i\psi$ iff $(\Ip(\gamma,\pi),r,m) \sat B_i \psi^B$. \eprf

In light of Theorem~\ref{genthm}, from this point on we work
with the larger system~$\Ip(\gamma,\pi)$ and use belief relative to
deviation-compatible ranking generators, instead of working 
with the system $\Irep(P,\gamma,\pi)$ and using knowledge.

By having both ranking generators and order generators in our framework,
we can handle both belief and counterfactual reasoning. 
Thus, for example, we can write $B_3(\doact(1,\sfa) \RCond \phi)$ to 
represent agent 3's belief that if agent 1 were to perform
action~$\sfa$ in the next round, then~$\phi$ would hold. 
We can further write 
$B_3(\doact(1,\sfa) \RCond \phi)~~\RCond~~\psi$ to state that were
it the case that agent~3 had the above belief, then in fact~$\psi$
would hold. 
Arbitrary nesting of belief and counterfactuals 
is allowed.
To take advantage of the expressive features of the framework, we now
define the analogue of knowledge-base programs, 
to allow for belief and counterfactuals. 

A {\em counterfactual belief-based program} (or {\em cbb program} for short) 
has  the same form as a knowledge-based program, except that the underlying 
logical language for the formulas appearing in tests is now~$\Lbc$ instead 
of~$\Lk$, and all tests in the program text~$\Pg_i$ for agent~$i$ are formulas
of the form~$B_i\psi$ or~$\neg B_i\psi$. 
As with knowledge-based programs, we are interested in when a protocol~$P$ 
{\em implements} a cbb program $\Pgcb$.
Again, the idea is that the protocol should act according to the high-level 
program, when the tests are evaluated relative to the counterfactual 
belief-based system corresponding to~$P$. 
To make this precise, given an extended system $\cJ=(\I,\lta,\rank)$
and a cbb program $\Pgcb$, let $\Pgcb^\cJ$
denote the protocol derived from $\Pgcb$ 
by using $\cJ$ to evaluate the belief tests.  That is, a
test in $\Pgcb$ such as $B_i\phi$ holds at a point $(r,m)$
relative to $\cJ$ if $\phi$ holds at all points $(r',m')$ in
$(\I,\kappa)$ such that $r'_i(m') =r_i(m)$ and $\rank(r') = 
\mini^\rank(r,m)$. 
Define an {\em extended context\/} to be a tuple
$(\gamma,\pi, \ogen,\sigma)$, where $(\gamma,\pi)$ is an interpreted context, 
$\ogen$ is an ordering generator for $\Rp(\gamma)$, 
and~$\sigma$ is a deviation-compatible ranking generator for $\gamma$.
An extended system $(\I,\lta,\rank)$ {\em represents\/}
the belief-based program
$\Pgcb$ in extended context $(\gamma,\pi,\ogen,\rgen)$ if 
(a) $\I = \Ip(\gamma,\pi)$, 
(b) $\lta=\ogen(\Pgcb^{(\I,\lta,\rank)})$, and 
(c) $\rank = \sigma(\Pgcb^{(\I,\lta,\rank)})$.
A protocol $P$ {\em implements\/} $\Pgcb$ in $(\gamma,\pi,\ogen,\sigma)$
if $P = \Pgcb^{(\Ip(\gamma,\pi),\ogen(P),\sigma(P))}$.
Protocol
$P$ {\em de facto implements\/} $\Pgcb$ in
$(\gamma,\pi,\sigma)$ if $P \approx_\gamma
\Pgcb^{(\Ip(\gamma,\pi),\ogen(P),\sigma(P))}$.

There is   a close connection 
between the notions of implementation for knowledge-based programs 
and implementation for cbb programs using
deviation-compatible rankings. 
Given a knowledge-based program~$\Pgkb$, we denote by $\Pgkb^B$ 
the program that results from replacing every knowledge operator~$K_i$ 
appearing in~$\Pkb$ to~$B_i$, for all agents $i=1,\ldots,n$. (This is, 
in particular, a cbb programs with no counterfactual operators.) 

\thm \label{b-imp}
Let $\Pgkb$ be a knowledge-based program and let~$\rgen$ be a 
deviation-compatible ranking generator for $\gamma$. 
Moreover, let~$\ogen$ be an arbitrary ordering generator for $\Rp(\gamma)$. 
A protocol $P$ de facto implements $\Pgkb$ in $(\gamma,\pi)$ 
if and only if 
$P$ de facto implements $\Pgkb^B$ in 
$(\gamma,\pi,\ogen,\rgen)$.
\ethm

\prf 
Since $\rgen$ is deviation compatible, by Theorem~\ref{genthm}, 
for all points $(r,m)$ of $\Rrep(P,\gamma)$, we have that
$(\Irep(P,\gamma,\pi),r,m)\sat\phi$ iff
$(\Ip(\gamma,\pi),\ogen(P),\rgen(P),r,m)\sat\phi^B.$  
Let $\Pgcb=\Pgkb^B$ and let $\cJ(P)=(\Ip(\gamma,\pi),\ogen(P),\rgen(P))$.
Then
\begin{equation}\label{eq1}
\mbox{$(\Pgkb)_i^{\Irep(P,\gamma,\pi)}(r_i(m))  =
(\Pgcb)_i^{\cJ(P)}(r_i(m))$ 
whenever $r \in \Rrep(P,\gamma)$.}  
\end{equation}
Now suppose that $P$ de facto implements 
$\Pgkb$.  By definition, $P \approx_\gamma \Pgkb^{\Irep(P,\gamma,\pi)}$.
Thus, 
the only global states that arise when running 
$\Pgkb^{\Irep(P,\gamma,\pi)}$ 
are those of the form 
$r(m)$ for some $r \in \Rrep(P,\gamma)$.  
It easily follows from (\ref{eq1}) that
$\I(\Pgkb^{\Irep(P,\gamma,\pi)},\gamma,\pi) =
\I(\Pgcb^{\cJ(P)},\gamma,\pi)$.  
Thus, $P$ de facto implements $\Pgcb$ as well.  
The argument in the other direction is 
analogous. \eprf
Theorem~\ref{b-imp} shows that a protocol $P$ de facto implements a
knowledge-based program iff $P$ de facto implements the corresponding
belief-based program.  Thus, 
by using deviation-compatible rankings, 
cbb programs can essentially emulate knowledge-based programs.
The move to cbb programs as defined here 
thus provides what may be considered a conservative extension 
of the knowledge-based framework: it allows us to treat
beliefs {\em and} counterfactuals, while being able to handle 
everything that the old theory gave us without changing the results.
\subsection{Analysis of the Bit-Transmission Problem}
Recall the program~$\Pgbt''_S$ from the introduction: 
{\bf if} $K_S(\recbit)$ {\bf then} \noop~{\bf else} {\sf sendbit}.
With this program, $S$ keeps sending the bit until it knows that
$R$ has received the bit. 
As discussed in the introduction, it would be even more efficient
for $S$ to stop sending the bit once it knows that {\em eventually\/} $R$ will
receive it.  As we saw, replacing $K_S(\recbit)$ by 
$K_S(\Diamond \recbit)$  leads to problems. We can deal with these problems
by using counterfactuals (and, thus, belief rather than knowledge), as
in the cbb program  $\PgbtR_S$ from the introduction: 
$$\mbox{{\bf if} $B_S (\doact(S,\noop) \RCond \Diamond \recbit)$
{\bf then} \noop~{\bf else} {\sf sendbit}.}$$
This program says that~$S$ should send the bit unless it believes that 
{\em even if it would not send the bit in the current round}, $R$ would
eventually receive the bit. 
Similarly, the program $\PgbtB_S$ says that $S$ should send the bit
unless it believes that $R$ would eventually correctly believe its value:
\[\mbox{{\bf if} $B_S (\doact(S,\noop) \RCond \Diamond B_R(\bit))$
{\bf then} \noop~{\bf else} {\sf sendbit}.}\]
(Recall that $B_R(\bit)$ is 
short for $(\bit = 0 \land B_R(\bit=0))
\lor (\bit=1 \land B_R(\bit = 1))$.)

Let $\PgbtR=(\PgbtR_S,\NOOP_R)$ and, similarly, let 
$\PgbtB=(\PgbtB_S,\NOOP_R)$.
We now consider the implementations of $\PgbtR$ 
and $\PgbtB$
in three different
contexts:
\begin{itemize}
\item 
$\gamma_1$, 
in which messages are guaranteed 
to be 
delivered within five rounds;%
\footnote{There is nothing special about five rounds here; another other
fixed number would do for the purposes of this example.}
\item 
$\gamma_2$, 
in which messages are guaranteed to arrive eventually,
but there is no upper bound on message delivery time;
and
\item 
$\gamma_3$, 
in which a message that is sent infinitely often is
guaranteed to arrive,
but there is no upper bound on message delivery time.
(Nothing can be said about a message sent only
finitely often; this is a standard type of {\em fairness\/} assumed in
the literature \cite{Francez86}.) 
\end{itemize}
In all contexts that we consider, messages cannot be reordered or duplicated. 
Moreover, a message can be delivered only if it was previously sent. 
We assume for now that we are working in synchronous systems,
so that processes can keep track of the round number.  (Indeed, we
cannot really make sense out of messages being delivered in five rounds
in asynchronous systems.)  At the end of this section we briefly comment
on how our results can be modified to apply to asynchronous systems.
We now describe these contexts more formally. 

In~$\gamma_1=(P_e^1,\Gz^1,\tau^1,\Psi^1)$, 
an agent can perform one of two actions: 
$\noop$ and $\sendbit$, with the obvious outcome. 
The local state of~$S$ consists of three components: 
(a) a Boolean variable {\it bit} that is fixed throughout the run, 
(b) a clock value, 
encoded in the variable $\clock$, 
which is always equal to the round number; at a
point $(r,m)$ the clock value is~$m$, and 
(c) the {\em message history}, which is the sequence of
messages that~$S$ has sent and received, each marked by time at which
it was sent or received. 
The local state of the receiver~$R$ consists of 
the clock value and~$R$'s message history. 
Assume that the set~$\Gz^1$ of initial states in~$\gamma_1$ consists of
two states---one in which $\bit=0$ and one in which  $\bit=1$.
In both states the clock values are~$0$ and message histories are empty.
In this context, 
messages are
guaranteed to be delivered within at most five rounds. 
The environment can perform the action of delivering a message.
Its protocol $P_e^1$ consists of deciding when messages are delivered,
subject to this constraint.
Since the environment's state keeps track of all actions performed,
it can be determined from the state
which messages are in transit and how long they have been in transit.
$\Psi^1$ makes no restrictions: all runs are considered admissible. 

The context $\gamma_2=(P_e^2,\Gz^2,\tau^2,\Psi^2)$ is a variant 
of~$\gamma_1$ with asynchronous communication. $\Gz^2=\Gz^1$, and 
the local states of~$S$ and~$R$ are the same as in~$\gamma_1$. 
Every message sent is guaranteed to be delivered, but there
is no bound on the time it will spend in transit. Thus, the environment's
state again keeps track of the messages in transit, while the environment's 
protocol $P_e^2$ decides at each point (nondeterministically) which,
if any, of the messages in transit should be delivered in the current round. 
The constraint that messages are guaranteed to eventually be delivered
is captured by the admissibility constraint~$\Psi^2$;
the set $\Psi^2$ consists of the 
runs in which every message sent is eventually delivered. 

The only difference between $\gamma_3 = (P_e^3,\Gz^3,\tau^3,\Psi^3)$ and 
$\gamma_2$ is 
that the admissibility condition $\Psi^3$ is more liberal than (i.e., is
a superset of) $\Psi^2$.  The set $\Psi^3$
consists of all runs~$r$ that are fair in the sense that,
for every time~$m$, if a given message $\mu$ is sent 
infinitely often in~$r$ after time~$m$, 
then at least one of the copies of~$\mu$ sent after time~$m$ is delivered. 

We define three sets of extended contexts, extending $\gamma_i$, $i = 1,
2, 3$.  Let $EC_i$ consist of all contexts of the form
$(\gamma_i,\pi,\ogen,\rgen)$, $i = 1, 2, 3$, 
where 
$\pi$ interprets the propositions $\bit=0$ and $\bit=1$
in the natural way, 
$\ogen$ respects protocols, and $\rgen$  is deviation compatible.
We claim that both $\PgbtR$ and $\PgbtB$ solve the bit-transmission problem 
in every extended context in $EC_i$, $i = 1, 2, 3$.  But what does it
mean for a protocol to ``solve'' the bit-transmission problem?  
To make this precise, we need to specify the problem.  In the case of
the bit-transmission problem, the specification is simple: we want the
receiver to eventually know the bit.  Thus, we say that a cbb-program
$\Pg$ solves the bit-transmission problem in extended context 
$\zeta = (\gamma,\pi,\ogen,\rgen)$ if,
for every protocol $P$ that de facto implements~$\Pg$, we have that
$(\Jp(P,\zeta),r,0)\sat \Diamond B_R({\bit})$ for
every run $r\in\Rrep(P,\gamma)$.
Notice that using belief here is safe, because we 
are requiring only that the belief hold in runs of $P$.  
Lemma~\ref{knowledge-axiom} guarantees that, in these runs (which all
have rank 0), the beliefs are true.

\thm\label{safe} Both $\PgbtR$ and $\PgbtB$ solve the bit-transmission
problem in all the extended contexts $EC_1 \union EC_2 \union EC_3$.
\ethm

\prf 
Let $\zeta= (\gamma,\pi,\ogen,\rgen)$ be a context in 
$EC_1 \union EC_2 \union EC_3$ and assume that~$P$ de facto 
implements~$\PgbtR$ 
or $\PgbtB$
in~$\zeta$. Let~$\cJ=(\cI(P,\gamma),\ogen(P),\rgen(P))$
and let $r\in\Rrep(P,\gamma)$ be a run of~$P$ in~$\gamma$.
We first consider the case that $P$ implements $\PgbtR$; the argument in
the case that $P$ implements $\PgbtB$ is even easier, and is sketched
afterwards.
There are two cases: 
\begin{itemize}
\item[(a)] Suppose that
$(\cJ,r,m)\sat B_S (\doact(S,\noop) \RCond \Diamond \recbit)$
for some $m > 0$.
Since~$P$ de facto implements~$\PgbtR$,
$S$ performs $\noop$ in round~$m+1$ of~$r$.
Thus, we have that
$(\cJ,r,m)\sat  \doact(S,\noop)$.
Since $\sigma(P)$ is deviation compatible and $r \in \Rrep(P, \gamma)$,  
it follows that 
$(\cJ,r,m)\sat  \doact(S,\noop) \RCond \Diamond \recbit$. 
Since $\ogen$ respects protocols, 
$(r,m)\in \closest(\intension{\doact(S,\noop)},(r,m),\cJ)$.
It now follows from the semantics of $\RCond$ 
that $(\cJ,r,m)\sat \Diamond \recbit$. 
Since $P$ de facto implements $\PgbtR$, if $S$ sends a value in a run
$r'$ of $P$, $S$ is actually sending the bit.  Since $\sigma(P)$ is
deviation compatible, it follows that in every run $r'$ of $P$, we have
that $(\cJ, r', m') \sat \recbit \rimp B_R (\bit)$, since all the points in
${\sf min}_R(r',m')$ are points on runs of $P$.  Thus, $(\cJ,r,m) 
\sat B_R(\bit)$.  
\item[(b)] Suppose that 
$(\cJ,r,m)\not\sat B_S (\doact(S,\noop) \RCond 
\Diamond \recbit)$
for all~$m\ge 0$. Since $P$ de facto implements~$\PgbtR$, 
it follows 
that~$S$ sends the bit in every round of~$r$. (In particular, the bit is 
sent by~$S$ 
infinitely often.) 
All three contexts under consideration have the property that a message sent 
infinitely often is guaranteed to be delivered. 
Thus, at some time~$m'\ge 0$ in~$r$, 
the receiver 
will receive the bit;  that is, $(\cJ,r,m') \sat \recbit$ for some $m' >
0$.  %
we have by 
Then, just as in part (a), it follows that $(\cJ,r,m') \sat B_R (\bit)$,
and hence
that $(\cJ,r,0)\sat \Diamond B_R(\bit)$.
\end{itemize}

The argument is almost identical (and somewhat simpler) if $P$
implements $\PgbtB$.  Now we split into two cases according to whether
there is some $m$ such that $(\cJ,r,m)\sat B_S (\doact(S,\noop) \RCond
\Diamond B_R(\bit))$.  Using the same arguments as above (but skipping
the argument that $\cJ \sat \recbit \rimp B_R(\bit)$) we get that,
in both cases, $(\cJ,r,0)\sat \Diamond B_R(\bit)$.
\eprf

Theorem~\ref{safe}, while useful, does not give us all we want.
In particular, it shows neither that $\PgbtR$ or $\PgbtB$ is
implementable nor 
that $S$ sends relatively few messages according to any
protocol that implements $\PgbtR$ 
or $\PgbtB$
(which, after all, was the goal of using
counterfactuals in this setting).  
In fact, as we now show, both $\PgbtR$ and~$\PgbtB$ are implementable
in all three sets of contexts, and their implementations are as
message-efficient as possible.  We consider each of $EC_1$, $EC_2$,and
$EC_3$ in turn.

Intuitively, in order to solve the bit-transmission problem 
in a context in which messages are always delivered, sending 
the bit only once in any given run should suffice. 
Consider the collection of protocols 
$P^1(k,m)=(P_S^1(k,m),\NOOP_R)$ for $k,m\in{\bf N}$, where 
$P_S^1(k,m)$ is described by the program
$$\mbox{{\bf if}  ($\clock=k$ and $\bit=0$) or ($\clock=m$ and $\bit=1$)
{\bf then} $\sendbit$ {\bf else} \noop}.$$
In these protocols, the sender~$S$ sends its bit 
at time~$k$ if
the bit value is~$0$, and at time~$m$ if it is~$1$. 
We now show that all protocols of the form $P^1(k,m)$ implement 
$\PgbtR$ in all contexts in $EC_1$:
\lem
\label{pbR-imp}
The protocol $P^1(k,m)$ de facto implements~$\PgbtR$ in every extended 
context in $EC_1$. 
\elem
\prf
Fix $k$, $m$, and a context $\zeta = (\gamma_1, \pi,  \ogen, \rank) 
\in EC_1$.  We want to show that $P^1(k,m) \approx_{\gamma_1}
(\PgbtR)^{\cJ(k,m)}$, where $\cJ(k,m) =
(\Ip(\gamma_1,\pi_1),\ogen(P^1(k,m)),\rgen(P^1(k,m)))$. 
We can characterize a run consistent with $P^1(k,m)$ by the value of
$\bit$ and when the one message sent by $S$ is received.  
Let $r_{b,n}$ be the run where $\bit = b$ and the message is received at
time $n$ (clearly $k+5\ge n > k$ if $b=0$ and $m+5\ge n > m$ if $b=1$).
Clearly
the formula $\recbit$ holds in run $r_{b,n}$ from time~$n$ on.
Thus, 
$\Diamond \recbit$ holds at every point in every run consistent with
$P^1(k,m)$ in the system $\cJ(k,m)$.  Note that the runs $r_{b,n}$ are
precisely those of rank 0 in $\cJ(k,m)$.

We now show that a run $r$ is consistent with 
$(\PgbtR)^{\cJ(k,m)}$ in $\gamma_1$ iff $r = r_{b,n}$ for $b \in \{0,1\}$ 
and a value of~$n$ satisfying 
$k+5\ge n > k$ if $b=0$ and $m+5\ge n > m$ if $b=1$.
So suppose that $r$ is consistent with $(\PgbtR)^{\cJ(k,m)}$ and the value of
the bit in $r$ is 0.  It suffices to 
show that $S$ sends exactly one message in~$r$, and that happens at time~$k$.
If $n' \ne k$, then clearly 
$(\cJ(k,m), r, n') \sat (S,\noop) \RCond \Diamond \recbit$, 
since the closest point to $(r,n')$ where $\doact(S,\noop)$ holds is
$(r,n')$ itself.  On the other hand, if $n' = k$,
then $\closest(\intension{\doact(S,\noop)},(r,n'),\cJ(k,m)) =
\{(r'_0,n')\}$, where $r'_0$ is the run where $S$ never sends any messages 
and the initial bit is $0$. In this case, the 
properties of~$\gamma_1$ guarantee that no message is ever received by~$R$
in~$r'$, and $\Diamond\recbit$ does not hold at $(r',k)$. 
It follows that the test 
$B_S (\doact(S,\noop) \RCond \Diamond \recbit)$ fails at $(r,k)$, and~$r$ 
is consistent with $\PgbtR$ if and only if the action $sendbit$ is performed 
in round~$k+1$ of~$r$. 
Hence, $r$ is one of the runs $r_{0,n}$ with $k+5\ge n>k$. 
A completely analogous treatment applies if $\bit=1$ in~$r$. 
We thus have that exactly the runs $r_{b,n}$ described are consistent
with $(\PgbtR)^{\cJ(k,m)}$ in~$\gamma_1$, and hence 
$P^1(k,m)$ de facto implements~$\PgbtR$ in every extended 
context in $EC_1$, as desired.
\eprf

In the context~$\gamma_1$, there is a fixed bound on message delivery time. 
As a result, we might hope to save on message delivery in some 
cases.
Suppose that we use a one-sided protocol, that sends the bit only if
$\bit=0$. Then the receiver should be able to 
conclude that the value of the bit is~1 if a message stating the bit is~0 
does not arrive within the specified time bounds. 
More generally, define the collection of protocols 
$P^2(k,b)=(P_S^2(k,b),\NOOP_R)$ 
for $b\in\{0,1\}$ and $k\in{\bf N}$, 
where $P_S^2(k,b)$ is the protocol implementing the program
\[\mbox{{\bf if}  $\clock=k$  and $\bit=b$
{\bf then} $\sendbit$ {\bf else} \noop}.\]
According to
$P_S^2(k,b)$, the sender $S$ sends a message only in runs where 
the bit is $b$; if the bit is $1-b$, it sends no messages.  Moreover, in
runs where the bit is $b$, $S$ sends only one message, at 
time $k$.
This type of optimization (sending a message only for one of the two bit
values) was used in the message-optimal protocols of \cite{HadHal};
it can be used in synchronous systems in which there is an upper bound on the
message delivery time, as in contexts in $EC_1$.

It is easy to verify that $P^2(k,b)$ does not implement $\PgbtR$:
Intuitively, in a run~$r$ of $P^2(k,b)$ with $\bit=1-b$, 
the sender~$S$ never sends the bit, and hence $\Diamond\recbit$ does not hold. 
Since $S$ follows $P^2(k,b)$ in~$r$, the formula
$\doact(S,\noop)$ holds at time~0 in~$r$. 
It follows that in evaluating the test 
$B_S (\doact(S,\noop) \RCond \Diamond \recbit)$ the closest point to $(r,0)$ 
is $(r,0)$ itself. Because $\Diamond\recbit$ does not hold at that point, 
the test fails, and according to~$\PgbtR$ the sender~$S$ should 
perform $\sendbit$. Since, in fact, $S$ does not perform $\sendbit$ 
at $(r,0)$, and~$r$ is a run of $P^2(k,b)$, 
we conclude that  $P^2(k,b)$ does not implement $\PgbtR$.
However, as we now show, $P^2(k,b)$
does implement the more sophisticated program~$\PgbtB$:

\lem\label{zeta-char1}
Every instance of $P^2(k,b)$ de facto implements 
$\PgbtB$ 
in every context
in $EC_1$.
\elem

\prf
Fix $k$, $b$, and a context $\zeta = (\gamma_1, \pi, \ogen,\rgen)
 \in EC_1$. 
  We want to show that $P^2(k,b) \approx_{\gamma_1}
(\PgbtB)^{\cJ(k,b)}$, 
where $\cJ(k,b) =
(\Ip(\gamma_1,\pi),\ogen(P^2(k,b)),\rgen(P^2(k,b)))$. 
Note that there are exactly 
six
runs consistent with $P^2(k,b)$ in context $\gamma_1$: 
five runs $r_b^m$, $m=k+1,\ldots,k+5$, where the value of the bit is~$b$, 
the message is sent in 
round~$k+1$ and 
it arrives in round~$m$; the sixth run is 
$r_{1-b}$, where the value of the bit is $1-b$ and no message is sent.
It is easy to check that in the extended system $\cJ(k,b)$, 
the formula 
$\bit=b\land B_R(\bit = b)$ holds in runs $r_b^m$ from 
time~$m$ on, while in run $r_{1-b}$ 
the formula $\bit=1-b\land B_R(\bit=1-b)$ holds from 
time $k+5$ on. Thus, $\Diamond B_R(\bit)$
holds at every point in the six runs in $\Rrep(P^2(k,b),\gamma_1)$.
Note that these 
six runs are exactly the runs of rank~0.

We now show that~$r$ is consistent with 
$(\PgbtB)^{\cJ(k,b)}$ iff $r \in \Rrep(P^2(k,b),\gamma_1)$.
We consider two cases, according to the values 
of the bit in~$r$.
First suppose that $\bit=1-b$ in the run~$r$.
We prove by induction on~$m'\ge 0$ that 
(a) if $r$ is consistent with $\Pgbt_S^{\cJ(k,b)}$ then 
(i) $r(m')=r_{1-b}(m')$ and
(ii) $(\PgbtB)_S^{\cJ(k,b)}(r_S(m')) = \noop$, and 
(b) $r_{1-b}$ is consistent with $(\PgbtB)_S^{\cJ(k,b)}$ up to time~$m'$.
For the base case, 
observe that $r(0)=r_{1-b}(0)$ because there is only one initial 
state in~$\gamma_1$ with $\bit=1-b$.  Clearly  $r_{1-b}$ is consistent
with $(\PgbtB)_S^{\cJ(k,b)}$ up to time~$0$.  
Thus, parts (a)(i) and (b) hold.  For part (a)(ii), 
to see that 
$(\PgbtB)_S^{\cJ(k,b)}(r_S(0)) = \noop$, 
it suffices to show that
$(\cJ(k,b),r,0) \sat B_S(\doact(S,\noop)\RCond\Diamond B_R(\bit))$. 
Since $\rgen_1$ is deviation compatible
and $S$ knows that $\bit= 1-b$, 
it follows that ${\sf
min}_S^{\rgen(P^2(k,b))}(r,0)=\{(r_{1-b},0)\}$.
Thus, it suffices to show that 
$(\cJ(k,b),r_{1-b},0) \sat \doact(S,\noop)\RCond\Diamond B_R(\bit)$. 
But this is immediate from the fact that 
$(\cJ(k,b),r_{1-b},0) \sat \doact(S,\noop)$ and, as observed earlier, 
that $(\cJ(k,b),r_{1-b},0) \sat \Diamond B_R(\bit)$.

For the inductive step in the case $\bit=1-b$, assume 
that the inductive claim holds for time~$m' \ge 0$. 
We want to show that it holds at time~$m'+1$.  Part (a)(i) and (b) are
immediate from the inductive hypothesis.
The argument for part (a)(ii) is the same as in the base case.  
This completes the inductive argument.  It follows immediately from the
induction that $r_{1-b}$ is consistent with $\Pgbt_S^{\cJ(k,b)}$ and
that if $r$ is consistent with $\Pgbt_S^{\cJ(k,b)}$ and $\bit = 1-b$ in
$r$, then $r = r_{1-b}$.  
Now consider the case where $\bit = b$ in~$r$.
Define $\bruns$ to be the set $\{r_{k+1},r_{k+2}\ldots,r_{k+5}\}$,
and $\bpoints(m')$ to be $\{(r_{k+1},m'),(r_{k+2},m'),\ldots,(r_{k+5},m')\}$.
We 
show by induction on $m' \ge 0$ that 
if $r$ is consistent with $(\PgbtB)^{\cJ(k,b)}$, then 
\begin{itemize}
\item[(a)] $r(m') \in \bpoints(m')$, 
\item[(b)] 
$(\PgbtB_S)^{\cJ(k,b)}
(r_S(m')) = \left\{
\begin{array}{ll}
\noop &\mbox{if $m'\ne k$}\\
{\sf sendbit} &\mbox{if $m' = k$,}
\end{array}
\right.
$
\item[(c)] at least one run in $\bruns$ agrees with $r$ up to
time $m'$; 
moreover, 
if $m' \ge k+5$, then exactly one run in $\bruns$ 
agrees with $r$ up to time $m'$.
\end{itemize}
For the base case, it is again immediate that $r(0) \in \bpoints(0)$ and
that all runs in $\bruns$ agree with $r$ up to time 0.
To see that part (b) holds, first
note that ${\sf min}_S^{\rgen(P^2(k,b))}(r,0)=\{(r^{k'},0)\stc k' = 1,
\ldots, 5\}$.  There are now two cases: if $k = 0$
(so that $S$ sends a message in round 1 of all the runs in $\bruns$),
then we must show that $(\cJ(k,b),r,0)\sat
\neg B_S(\doact(S,\noop)\RCond\Diamond B_R(\bit))$, so that 
$\Pgbt_S^{\cJ(k,b)}(r_S(0)) = {\sf sendbit}$.  
Note that, if $k=0$, then
$\closest(\intension{\doact(S,\noop)},(r^{k'},0),\cJ(k,b)) = \{r^*\}$
for $k' = 1, \ldots, 5$, where~$r^*$ 
is the run where $\bit = b$ and no messages are ever sent by~$S$ or~$R$.  
Thus, it suffices 
to show that $(\cJ(k,b),r*,0)\sat \neg \Diamond B_R(\bit)$.  It is easy
to see that, since $\rgen_1$ is deviation compatible, we must have
$(r_{1-b},m) \in  {\sf min}_R^{\rgen(P^2(k,b))}(r^*,m)$, 
for all $m \ge 0$.  
Thus, $(\cJ,r^*,m) \not\sat \bit=1-b \land B_R(\bit = 1-b)$ for all 
$m\ge 0$, 
and hence $(\cJ,r^*,m') \sat \neg \Diamond B_R(\bit)$ for all $m'\ge 0$, 
as desired.  On the other hand, if $k > 0$, we must show that 
$(\cJ(k,b),r,0)\sat B_S(\doact(S,\noop)\RCond\Diamond B_R(\bit))$.  
Note that if $k > 0$, then
$\closest(\intension{\doact(S,\noop)},(r^{k'},0),\cJ(k,b)) = 
\{r^{k'}\}$, for $k' = 1, \ldots, 5$.   Since $(\cJ(k,b),r^{k'},0)\sat
\doact(S,\noop) \land B_R(\bit))$, we are done.

The argument in the inductive step is almost identical, except 
that it 
now breaks into the cases $m' < k$, $m' = k$, $k < m' < k+5$, and $m' \ge
k+5$.  We leave details to the reader. 

Finally, we must show that each run $r \in \bruns$ is consistent with 
$(\PgbtB_S)^{\cJ(k,b)}$.  
We proceed by induction on $m'$ to show that
$r$ is consistent with 
$(\PgbtB_S)^{\cJ(k,b)}$
up to time $m'$.  This
involves proving part (b) of the induction above for each $r \in
\bruns$.  The proof is similar to that above, 
and left to the reader.
\eprf

The preceding discussion has shown that $P^2(k,b)$ implements $\PgbtB$,
but not $\PgbtR$, in contexts in $EC_1$.  Lemma~\ref{pbR-imp} shows that
$P^1(k,m)$ implements $\PgbtR$ in contexts in $EC_1$.  An obvious question is
whether $P^1(k,m)$ implements $\PgbtB$ in contexts in $EC_1$.  
We now show 
that if $k \ne m$, then $P^1(k,m)$ does {\em not\/} implement
$\PgbtB$; if $k = m$, then whether $P^1(k,m)$ implements $\PgbtB$ 
depends on what the receiver believes in runs where he does not receive a
message.  Since there is no run of $P^1(k,m)$ where the receiver
receives no messages, this is not determined by just assuming that we
have a deviation-compatible ranking generator.  Given a ranking
$\kappa$, let $\rank(n,b)$ be the rank of the run with least rank where
(a) the receiver does not receive any messages up to
and including time $n$ and (b) the bit has value $b$.  
We say that a ranking $\rank$ is {\em biased\/} 
if $\rank(n,0) \ne \rank(n,1)$ holds for at least one time instant~$n$. 
Note that if $\rank(n,i) < \rank(n,i\oplus 1)$ then, in the absence of
messages, $R$ will believe that the bit is $i$ at time $n$.

\lem\label{zeta-char21}
Let $\zeta = (\gamma_1, \pi, \ogen,\rgen)  \in EC_1$.
The protocol $P^1(k,m)$ de facto implements $\PgbtB$ in~$\zeta$ 
exactly if 
both
(a) $k=m$ and (b) $\rgen(P^1(k,k))$ is not biased.
\elem
\prf
Fix a context $\zeta = (\gamma_1, \pi, \ogen,\rgen) \in EC_1$.
As in the proof of Lemma~\ref{pbR-imp}, define
$\cJ(k,m) = (\Ip(\gamma_1,\pi),\ogen(P^1(k,m)),\rgen(P^1(k,m)))$
and the runs $r_{b,n}$. 

First suppose that $\rgen(P^1(k,k))$ is not biased.  We show that
$P^1(k,k)$ de facto implements $\PgbtB$ in~$\zeta$. 
By definition, in each of the ten
runs $r_{b,n}$ of rank 0 in the extended system $\cJ(k,k)$, $\recbit$
holds at the time $n$ when the receiver $R$ receives the bit.  Since $R$
receives the correct bit, it is easy to see that in fact
$(\cJ(k,k),r_{b,n},n) \sat 
B_R(\bit)$.   Thus, $\Diamond B_R(\bit)$
holds at every point in the ten runs of the form $r_{b,n}$ in the system
$(\PgbtB)^{\cJ(k,k)}$.  
Moreover, $(\cJ(k,k),r_{b,n},m) \sat
\doact(S,\noop)\RCond\Diamond B_R(\bit)$ for $m\ne k$.  Since 
the runs $r_{b,n}$ are
the runs of rank 0, it actually follows that 
$(\cJ(k,k),r_{b,n},m) \sat
B_S(\doact(S,\noop)\RCond\Diamond B_R(\bit))$ for $m \ne k$.  
We now show that
$(\cJ(k,k),r_{b,n},k) \sat \neg
B_S(\doact(S,\noop)\RCond\Diamond B_R(\bit))$.
Note that $\closest(\intension{\doact(S,\noop)},(r_{b,n},k),\cJ(k,k)) =
\{(r_b',k)\}$,  
where $r'_b$
is the run where the bit is~$b$ and $S$ sends no messages.
Suppose that $(\cJ(k,k),r_b',k) \sat \Diamond B_R(\bit = b)$.  
Thus, there is some $n \ge k$ such that $(\cJ(k,k),r_b',n) \sat 
B_R(\bit = b)$.   Then we must have $\rank(n,b) < \rank(n,b\oplus 1)$, so that
$\rank$ is biased, contradicting the assumption. 
Thus, $(\cJ(k,k),r_b',k) \sat \neg\Diamond B_R(\bit = 0)$, so 
$(\cJ(k,k),r_{b,n},k) \sat \neg
B_S(\doact(S,\noop)\RCond\Diamond B_R(\bit))$, as desired.  
In this case, by $(\PgbtB)^{\cJ(k,k)}$, the sender~$S$ should 
perform~$\sendbit$ at time~$k$. 
It follows that $r_{b,n}$ is consistent with $(\PgbtB)^{\cJ(k,k)}$.

We next show that if $r$ is consistent with 
$(\PgbtB)^{\cJ(k,k)}$, then $r \in \{r_{b,n}: b = 0, 1, \, n = k+1, \ldots,
k+5\}$.   
So suppose that the bit is 0 in $r$ and that $r$ is
consistent with $(\PgbtB)^{\cJ(k,k)}$.
Just as in the proof of Lemma~\ref{zeta-char1}, it is easy to show
by induction on $m$ 
that no messages are sent in $r$ at time  $m <k$:  It is easy to see that
$(\cJ(k,k),r,
m) \sat B_S(\doact(S,\noop)\RCond\Diamond B_R(\bit))$ for $k < m$, since
$(r,m) \sim_R (r_{b,n},m)$.  Just as in the case of $r_{b,n}$, we can
show that $(\cJ(k,k),r,k) \sat \neg
B_S(\doact(S,\noop)\RCond\Diamond B_R(\bit))$.  Thus, since $r$ is
consistent with $(\PgbtB)^{\cJ(k,k)}$, 
the sender 
$S$ sends a message at time $k$
in $r$.  It is easy to show that $S$ does not send the bit after time $k$;
we leave details to the reader.  
Thus, if $r$ is consistent with $(\PgbtB)^{\cJ(k,k)}$ then $S$ sends the
bit in $r$ at time $k$ (and does not send it at any other time), so $r$
is of the form $r_{b,n}$.

We next claim that if $k\ne m$ then 
$P^1(k,m)$ does not de facto implement $\PgbtB$ in~$\zeta$. 
Without loss of generality, suppose that $k<m$. 
By the properties of~$\gamma_1$, messages can take up to five time 
units to be delivered. Hence, there is a run of $P^1(k,m)$  
with $\bit=1$ in which the sender's message is not delivered by time~$m+4$. 
However, because $k<m$, there is no 
run with $\bit=0$ where no
message is delivered by time~$m+4$.
Because $\rgen$ is deviation compatible, it
follows that $\rank(m+4,1)=0<\rank(m+4,0)$. 
Thus,  $(\cJ(k,m), r_{1,m+4}, m+4) \sat B_R(\bit = 1)$, 
so $(\cJ(k,m), r_{1,m+j}, m) \sat B_S(\doact(S,\noop)\RCond\Diamond
B_R(\bit))$ for $j = 1, \ldots, 5$.  
Therefore, $S$ should not send the bit at time~$m$ according 
to $(\PgbtB)^{\cJ(k,m)}$
in runs where the bit is 1, showing
that $P^1(k,m)$  does not de facto implement $\PgbtB$. 

To complete the proof of the lemma, we need to show  that 
if $\kappa=\rgen(P^1(k,k))$ is biased, then 
$P^1(k,k)$ does not implement~$\PgbtB$ in $\zeta$.  
So suppose that~$\kappa=\rgen(P^1(k,k))$ is biased. 
Since~$\kappa$ is biased, 
there is an~$n$ for which $\kappa(n,0)\ne\kappa(n,1)$. 
Without loss of generality, assume that $\kappa(n,0) < \kappa(n,1)$.  
We must have $n > k$, since $\kappa(\ell,0)=\kappa(\ell,1)=0$ for all 
\mbox{$\ell \le k$},
because  in all runs consistent with $P^1(k,k)$,
the receiver~$R$ receives 
no messages up to time~$\ell$.  
It follows that $(\cJ(k,k),r,k) \sat \Diamond B_R(\bit=0)$ 
for all runs~$r$ consistent with $P^1(k,k)$. 
Thus, $(\cJ(k,k), r_{0,k+j}, m) \sat B_S(\doact(S,\noop)\RCond\Diamond
B_R(\bit))$ for $j = 1, \ldots, 5$.  
It follows that, in runs where the bit is 0, $S$ should not send the bit
according to  
$(\PgbtB)^{\cJ(k,k)}$. 
This again establishes that $P^1(k,k)$ 
does not de facto implement $\PgbtB$. 
\eprf

Now consider the context~$\gamma_2$. Here there is no upper bound 
on message delivery times. As a result, $S$ must 
send~$R$ messages regardless of what bit value is. 

\lem\label{zeta-char2}
Every instance of $P^1(k,m)$ de facto implements 
both $\PgbtR$ and $\PgbtB$ in every context in $EC_2$. 
\elem
\prf
The proof for the case of $\PgbtR$ is identical to the proof 
given for contexts in $EC_1$ in Lemma~\ref{pbR-imp}. There are now 
infinitely many runs $r_{b,n}$ consistent with $P^1(k,m)$ rather than
ten runs, but  
the argument remains sound. We leave details to the reader. 

In the case of $\PgbtB$, the argument follows the same lines as the proof 
Lemma~\ref{pbR-imp}, 
except that the role of $\Diamond\recbit$ is now played by 
$\Diamond B_R(\bit)$. 
Fix $k$, $m$, and a context $\zeta = (\gamma_2, \pi, \ogen, \rgen) 
\in EC_2$.  We want to show that $P^1(k,m) \approx_{\gamma_2}
(\PgbtB)^{\cJ'(k,m)}$, where $\cJ'(k,m) =
(\Ip(\gamma_2,\pi),\ogen(P^1(k,m)),\rgen(P^1(k,m)))$. 
It is easy to check that in the extended system
$\cJ'(k,m)$, the formula 
$B_R(\bit = b)$ holds in run $r_{b,n}$ from time~$n$ on.
Thus, 
$\Diamond B_R(\bit)$ holds at every point in every run consistent with
$P^1(k,m)$ in the system $\cJ'(k,m)$.  Note that the runs $r_{b,n}$ are
precisely those of rank 0 in $\cJ'(k,m)$.
Finally, note that if $(r',n)$ is an arbitrary point in $\cJ'(k,m)$ with
$n > \max(k,m)$ and no messages are 
sent in $r'$ up to time $n$, then $(\cJ'(k,m), r', n) \sat \neg B_R (\bit
= 0) \land \neg B_R (\bit = 1)$, since there are runs consistent with
$P^1(k,m)$ where no messages arrive up to time $n$ and the bit can be
either 0 or 1; for example, $(r_{0,n+1},n) \sim_R (r',n)$ and 
$(r_{1,n+1},n) \sim_R (r',n)$.

We now show that a run $r$ is consistent with 
$(\PgbtB)^{\cJ'(k,m)}$ in
$\gamma_2$ iff $r = r_{b,n}$ for $b \in \{0,1\}$ and $n \ge 0$.
So suppose that $r$ is consistent with 
$(\PgbtB)^{\cJ'(k,m)}$ and the value of
the bit in $r$ is 0.  It suffices to 
show that $S$ sends exactly one message in~$r$, and that happens at time~$k$.
The argument is very similar to that in Lemma~\ref{zeta-char1}.
If $n < k$, then 
clearly 
$(\cJ'(k,m), r, n) \sat (S,\noop) \RCond \Diamond B_R(\bit)$, 
since the closest point to $(r,n)$ where $\doact(S,\noop)$ holds is
$(r,n)$ itself.  On the other hand, if $n = k$, 
then $\closest(\intension{\doact(S,\noop)},(r,n),\cJ'(k,m)) =
\{(r'_0,n)\}$, where $r'_0$ is the run where $S$ sends no
messages and the initial bit is $0$.  As observed earlier,
we have $(\cJ'(k,m),r'_b,n) \sat \Box(\neg B_R (\bit
= 0) \land \neg B_R (\bit = 1))$, so
$(\cJ'(k,m),r_{b,n},n) \sat 
\neg (\doact(S,\noop) \RCond \CorrectB_R(\bit))$.  
Thus, since $r$ is
consistent with 
$(\PgbtB)^{\cJ'(k,m)}$ in $\gamma_2$, $S$ sends its bit
at time $k$ in 
$r$.  Finally, if $n > k$, again we have
$\closest(\intension{\doact(S,\noop)},(r,n),\cJ'(k,m)) = \{(r,n)\}$
so, again, $S$ does not send a message at time $n$ in $r$.
Thus, $r$ has the form $r_{0,n'}$ for some~$n'$.  The same argument shows
that all runs of the form $r_{0,n'}$ are in fact consistent with
$(\PgbtB)^{\cJ'(k,m)}$.  
The argument if $b=1$ is identical (with $m$
replacing $k$ throughout).  \eprf

Finally, we consider the contexts in $EC_3$.
In this case, communication is such that if~$R$ sends no 
messages, then $S$ is guaranteed to have one of its messages delivered 
only in case it sends infinitely many message. 
This says that if we consider only protocols of the form
$(P_S,\NOOP_R)$, then $S$ must send infinitely many messages in every
run.  However, if a protocol sends infinitely many messages, then no
particular one is necessary; if $S$ does not send, say, the first
message, then it still sends infinitely many, and $R$ is guaranteed to
get a message.  This suggests that we will have difficulty finding a
protocol that implements $\PgbtR$ or $\PgbtB$.
The following proposition prevides further evidence of this.
If $I \subseteq \IN$ (the natural numbers), let
$P(I)=(P_S(I),\NOOP_R)$, 
where $P_S(I)$ is described by the program
$$\mbox{{\bf if}  $\clock\in I$ {\bf then} $\sendbit$ {\bf else}
\noop}.$$
Thus, with $P_S(I)$, the sender $S$ sends the bit at 
every time that appears in~$I$.
\pro
\label{procra}
No protocol of the form $P(I)$ de facto implements either $\PgbtR$ or 
$\PgbtB$ in any context in~$EC_3$.
\epro
\prf
We sketch the argument here and leave details to the reader. 
First suppose that $I$ is finite.  Let $r$ be a
run in 
$P(I)$ 
where none of the finitely many messages sent by
$S$ is received.  Let $n = \sup(I) + 1$.  Suppose that 
$(\gamma_3,\pi,\ogen,\rgen) \in EC_3$.  
Let $\cJ(I) = 
(\Ip(\gamma_3,\pi),\ogen(P(I)),\rgen(P(I)))$. 
Clearly, 
$\closest(\intension{\doact(S,\noop)},(r,n),\cJ(I)) = \{(r)\}$,
since $S$ performs the act $\noop$ at $(r,n)$.  However,
since $R$ never receives the bit in run $r$, and $\rgen(P(I))(r) = 0$,
it follows that $(\cJ(I),r,n) \sat \neg \Diamond \recbit$ and 
$(\cJ(I),r,n) \sat \neg B_R(bit)$.  Thus, according to both $\PgbtR$ and
$\PgbtB$, $S$ should send a message at $(r,n)$.  It follows that $P(I)$
does not implement $\PgbtR$ or $\PgbtB$.

Now suppose that $I$ is infinite.
The properties of~$\gamma_3$ ensure that 
$R$ does in fact receive the bit in every run of $P(I)$.
Moreover, it is easy to check that when the message is received, 
both $\recbit$ and $B_R(\bit)$ 
hold.
Hence, for any given clock time $m\in I$, 
the formulas $\doact(S,\noop) \RCond\Diamond\recbit$ and 
$\doact(S,\noop) \RCond\Diamond B_R(\bit)$ hold at time~$m$ in all runs of
the protocol. A straightforward argument shows that $\sendbit$ is 
neither compatible with  $\PgbtR$ nor with $\PgbtB$ at time~$m$.  
\eprf

Intuitively, Proposition~\ref{procra} is a form 
of
the ``procrastinator's paradox'': Any action 
that must be performed only eventually (e.g., washing the dishes) 
can always safely be postponed for one more day. Of course, using this 
argument inductively results in the action never being performed.

Despite Proposition~\ref{procra}, we now show that 
$\PgbtR$ and $\PgbtB$ are both implementable in all contexts in~$EC_3$. 
Let $P^\omega=(P^\omega_S,\NOOP_R)$, where $P^\omega_S$ is 
the protocol determined by the following program: 
$$\mbox{{\bf if} $\clock=0$ or $\sendbit$ was performed in 
the previous round, {\bf then} $\sf sendbit$ {\bf else} $\noop$}.$$
Since $S$'s local state contains both the current time and 
a record of the time at which it sent every previous message, 
it can perform the test in $P^\omega(S)$.
It is not too hard to see that $P^\omega$ is de facto equivalent 
to $P(\IN)$ in~$\gamma_3$---under normal circumstances 
the bit is sent in {\em each and every} round. 
The two protocols differ only in their counterfactual behavior. 
As a result, while $P(\IN)$ 
implements neither $\PgbtR$ nor $\PgbtB$, the protocol 
$\Po$ implements both.

\lem\label{zeta-char3}
$\Po$ de facto implements 
both $\PgbtR$ and $\PgbtB$ in every context in $EC_3$. 
\elem
\prf
We provide the proof for $\PgbtB$. The proof for $\PgbtR$ is 
similar, and left to the reader. 

Fix a context $\zeta_3 = (\gamma_3, \pi,  \ogen, \rgen) \in EC_3$.
We want to show that $\Po \approx_{\gamma_3}
(\PgbtB)^{\cJo}$, where $\cJo =
(\Ip(\gamma_3,\pi),\ogen(\Po),\rgen(\Po))$. 
Let $\Ro=\Rrep(\Po,\gamma_3)$. 
Note that, for every natural number~$k$, 
there are runs $r_{b,k}\in\Ro$ in which $\bit=b$ and no message that is 
sent by~$S$ in the first~$k$ rounds is ever delivered to~$R$. 
It follows that
if~$R$ has received no message by time~$m$ in run~$r$ of~$\Ro$, 
then $(\cJo,r,m)\sat\neg B_R(\bit)$.
We now prove by induction on~$k$ that a run~$r$ is consistent with 
$(\PgbtB)^{\cJo}$ in~$\gamma_3$ 
for~$k$ rounds exactly if $S$ has performed~$\sendbit$ in each of the 
first~$k$ rounds of~$r$. 
The base case for $k=0$ is vacuously true. 
For the inductive step, assume that the claim is true 
for
$k=\ell$.
Suppose 
that $r$ is consistent 
with $(\PgbtB)^{\cJo}$ for~$\ell+1$ rounds. 
By the induction hypothesis,
the sender~$S$ has performed~$\sendbit$ in each of the 
first~$\ell$ rounds. Since~$r$ is, by assumption, 
consistent with $(\PgbtB)^{\cJo}$ for~$\ell+1$ rounds, $S$
performs $\sendbit$ in round~$\ell+1$ of~$r$ 
exactly if 
$(\cJo,r,\ell)\sat \neg B_S (\doact(S,\noop) \RCond \Diamond B_R(\bit))$.
Let $\bit=b$ in~$r$. 
Moreover, 
$\rgen_3(\Po)(r)=0$ 
since $\rgen_3$ is deviation compatible. 
Clearly $(r,\ell)\sim_S(r_{b,\ell})$, where $r_{b,\ell} \in \Ro$ is the run
constructed earlier where none of the message sent by $S$ in the first
$\ell$ rounds arrive, since in both $r$ and $r_{b,\ell}$, the bit is the
same and $S$ sends a message in each of the first $\ell$ rounds.
Moreover, $\rgen(\Po)(r_{b,\ell})=0$, since $\rgen$ is deviation compatible 
and $r_{b,\ell}\in\Ro$. 
Thus, to show that $(\cJo,r,\ell)\sat \neg B_S (\doact(S,\noop) \RCond
\Diamond B_R(\bit))$, it suffices to show that 
$(\cJo,r_{b,\ell},\ell)\sat \neg (\doact(S,\noop) \RCond \Diamond B_R(\bit))$.
The points in 
$\closest(\intension{\doact(S,\noop)},(r,\ell),\cJo)$
have the form $(r',\ell)$ where $r'$ agrees with $r_{b,\ell}$ up to
and including  time~$\ell$, 
$S$ does nothing in round $\ell$ of $r'$,
and~$S$ follows~$\Po$ in all rounds after~$\ell$ in~$r'$. 
The key point here is that, by following $\Po$, $S$ sends no messages in
$r'$ after round $\ell$.
Consequently, in all runs appearing in this set of closest points, $S$ sends
a finite number of message (exactly~$\ell$, in fact). 
By the admissibility condition $\Psi^3$ of~$\gamma_3$, 
there is one run in this set, which we denote by~$\hat r$, 
in which $R$ receives no messages.
Note that $(\hat{r},n) \sim_R (r_{0,n},n)$ and $(\hat{r},n) \sim_R
(r_{1,n},n)$, since in all of $\hat{r}$, $r_{0,n}$ and $r_{1,n}$,
the receiver $R$ receives no messages up to time $n$.  Since both
$r_{0,n}$ and $r_{1,n}$ are in $\Ro$, it follows that they both have
rank 0.  Thus, $(\cJo,\hat{r},n) \sat \neg B_r(\bit)$.  That is,
$B_R(\bit)$ never holds in~$\hat r$. 
It follows that 
$(\cJo,r_{b,\ell},\ell)\sat \neg (\doact(S,\noop) \RCond \Diamond B_R(\bit))$, 
as needed. 
We can thus conclude that $r$ is consistent with 
$(\PgbtB)^{\cJo}$ in~$\gamma_3$ for~$\ell+1$ rounds exactly if 
$S$ performs $\sendbit$ in the first~$\ell+1$ rounds, and we are done. 
\eprf

Lemma~\ref{zeta-char3} shows one way of resolving the procrastinator paradox: 
If one decides that an action (e.g., washing the dishes) 
that is not performed 
now
will {\em never} be performed, 
then performing it becomes critical.
(We are ignoring the issue of how one can ``decide'' to use such
protocol.  In the context of distributed computing, we can just make this
the protocol; people are likely not to believe that this is truly the
protocol.)  
In any case, using such a protocol makes performing
the action consistent with the procrastinator's protocol of doing 
no more than what is absolutely necessary.

We can summarize our analysis of implementability of $\PgbtR$ and $\PgbtB$ 
by the following theorem:

\thm 
\label{imp-thm}
Both $\PgbtR$ and $\PgbtB$ are de facto implementable in every extended 
context in $EC_1 \union EC_2\union EC_3$
Moreover, if $P$ de facto implements 
$\PgbtR$ or $\PgbtB$ in a context $\zeta \in EC_1 \union EC_2$, 
then $S$ sends at most one message in
every run consistent with $P$ in $\zeta$.
\ethm

\prf 
The implementability claims follow from Lemmas~\ref{pbR-imp}, 
\ref{zeta-char1}, and \ref{zeta-char3}. 
We now prove that $S$ sends no more than one message in every run of a
protocol that de facto implements $\PgbtR$ or $\PgbtB$ in a context in
$EC_1 \union EC_2$.
Suppose that $P = (P_S, P_R)$ de facto implements $\Pgbt_R$ in
$\zeta = (\gamma, \pi, \ogen,\rgen) \in EC_1 \union EC_2$.
Further suppose, by way of contradiction, that there is a run
$r$ consistent with $P$ in $\gamma$ in which the sender sends more
than one message.  Suppose that the second message is sent at time~$k$,
and the value 
of the bit in $r$ is $b$.  Let $\cJ
=(\Ip(\gamma,\pi),\ogen(P),\rgen(P))$. 
Since $\gamma \in
\{\gamma_1,\gamma_2\}$, all messages are guaranteed to arrive eventually 
in the context $\gamma$.  Thus, it is easy to see that
$(\cJ,r,k) \sat B_S(\Diamond B_R(\bit = b))$.
It follows that
$(\cJ,r,k) \sat \doact(S,\noop) \RCond B_R(\bit)$.  Since $P$ is de facto
consistent with 
$\PgbtR$, this means that $S$ should not send a message at $(r,k)$.
This is a contradiction.  
\eprf

All the contexts we have considered are synchronous; the sender and
receiver know the time.  As we observed earlier, there is no analogue of
$\gamma_1$ in the asynchronous setting, since it does not make sense to say
that messages arrive in 5 rounds.  However, there are obvious analogues
of $\gamma_2$ and $\gamma_3$.  Moreover, if we assume that $S$'s local
state keeps track of how many times it has been scheduled and what it
did when it was scheduled, then 
the analogue of $P^2(k,m)$ implements both $\PgbtR$ and
$\PgbtB$ if messages are guaranteed to arrive (where now $P^2(k,m)$
means that if $\bit = 0$, then 
the $k$th time that $S$ is scheduled it performs $\sendbit$, while if
$\bit = 1$, then the $m$th time that $S$ is scheduled it performs
$\sendbit$).  Similarly, the analogue of $\Po$ implements both $\PgbtR$
and $\PgbtB$ 
in contexts that satisfy the fairness assumption (but any
finite number of messages may not arrive).  

\section{Discussion}
\label{discussion}

This paper presents a framework that facilitates high-level 
counterfactual reasoning about protocols. Indeed, it enables the 
design of well-defined protocols in which processes act based on 
their knowledge of counterfactual statements. 
This is of interest because, in many instances, 
the intuition behind the choice of a given course of action is best thought
of and described in terms of counterfactual reasoning. For example, 
it is sometimes most efficient for agents to stop exending resources
once they know that their goals will be achieved even if they stop.
Making this precise involves counterfactual reasoning; this agent must
consider what would happen were it to stop expending resources.

This paper should perhaps best be viewed as a ``proof of concept'';
the examples involving the bit-transmission program 
show that counterfactuals can play a useful role in knowledge-based
programs.  While we have used standard approaches to giving semantics
to belief and counterfactuals (adapated to the runs and systems
framework that we are using), these definitions give 
the 
user a large
number of degrees of freedom, in terms of choosing the ranking function
to define belief and the notion of closeness needed to define
counterfactuals.  While we have tried to suggest some reasonable choices
for how the ranking function and the notion of closeness 
are
defined, and
these choices certainly gave answers that matched 
our
intuitions in all the
context we considered for the bit-transmission problem, it
would be helpful to have a few more examples to test the reasonableness
of the choices.  
We are currently exploring the application of cbb programs for analyzing
message-efficient leader election in various topologies; we hope to
report on this in future work.

While we used the very simple problem of bit transmission as a vehicle
for introducing our framework for knowledge, belief, and
counterfactuals, we believe it should be useful for handling a much
broader class of distributed protocols. 
We gave an example of how counterfactual reasoning is useful in deciding
whether a message needs to be sent.  Similar issues arise,
for example, in deciding whether to perform a write action
on a shared-memory variable. 
Because our framework provides a concrete model
for understanding the interaction between belief and counterfactuals, 
and for defining the notion of ``closeness'' needed 
for interpreting
counterfactuals, it
should also be useful for illuminating some problems in philosophy and
game theory.  
The insight our analysis gave to the procrastinator's paradox is an 
example of how counterfactual programs can be related to issues in 
the philosophy of human behavior. 
We believe that, in particular, the framework will be
helpful in understanding some extensions of Nash equilibrium in game
theory.  
For example, as we saw in Lemma~\ref{zeta-char21},
whether a protocol de facto
implements a cbb program depends on the agent's beliefs.
This seems closely related to 
the notion of a {\em subjective equilibrium\/} in 
game theory \cite{KaL95}. 
We are currently working on drawing a formal connection between our framework 
notions of equilibrium in game theory. 
It would also be interesting to relate the notion of ``closeness''
defined in our framework to that given by the structural-equations model 
used by Pearl \citeyear{pearl:2k} (see also \cite{Hal20}).  The
structural-equations model also gives a concrete interpretation to
``closeness''; it does so in terms of mechanisms defined by equations.
It would be interesting to see if these mechanisms can be modeled as
protocols in a way that makes the definitions agree.

\bibliographystyle{chicago}
\bibliography{z,refs,joe}
\end{document}

%% file: bookdefn.tex
\newlength{\defitemindent}
     {\begin{list}{}
               {\labelsep=0ex
            \settowidth{\labelwidth}{\hspace{\defitemindent} #1 \hspace{0.5ex}}
            \leftmargin=\labelwidth
            \addtolength{\leftmargin}{\labelsep}
             }}
     {\end{list}}

           {\arraycolsep=0.14em
                      \begin{eqnarray}}
                      {\end{eqnarray}}
 
\newenvironment{Eqnarray*}%
           {\arraycolsep=0.14em
                      \begin{eqnarray*}}
                      {\end{eqnarray*}}

\newcommand{\YOU}{\nonscript\kern 0.2ex U \nonscript\kern-0.2ex}
\newtheorem{THEOREM}{Theorem}[section]
\newenvironment{theorem}{\begin{THEOREM}\ }%
                        {\end{THEOREM}}
\newtheorem{LEMMA}[THEOREM]{Lemma}
\newenvironment{lemma}{\begin{LEMMA}\ }%
                      {\end{LEMMA}}
\newtheorem{COROLLARY}[THEOREM]{Corollary}
\newenvironment{corollary}{\begin{COROLLARY}\ }%
                          {\end{COROLLARY}}
\newtheorem{PROPOSITION}[THEOREM]{Proposition}
\newenvironment{proposition}{\begin{PROPOSITION}\ }%
                            {\end{PROPOSITION}}
\newtheorem{DEFINITION}[THEOREM]{Definition}
\newenvironment{definition}{\begin{DEFINITION}\ \rm}%
                            {\end{DEFINITION}}
\newtheorem{CLAIM}[THEOREM]{Claim}
\newenvironment{claim}{\begin{CLAIM}\ \rm}%
                            {\end{CLAIM}}
\newtheorem{EXAMPLE}[THEOREM]{Example}
\newenvironment{example}{\begin{EXAMPLE}\ \rm}%
                            {\end{EXAMPLE}}
\newtheorem{REMARK}[THEOREM]{Remark}
\newenvironment{remark}{\begin{REMARK}\ \rm}%
                            {\end{REMARK}}
\newcommand{\thm}{\begin{theorem}}
\newcommand{\lem}{\begin{lemma}}
\newcommand{\pro}{\begin{proposition}}
\newcommand{\dfn}{\begin{definition}}
\newcommand{\rem}{\begin{remark}}
\newcommand{\xam}{\begin{example}}
\newcommand{\cor}{\begin{corollary}}
\newcommand{\prf}{\noindent{\bf Proof} \hspace{0.677em}}
\newcommand{\ethm}{\end{theorem}}
\newcommand{\elem}{\end{lemma}}
\newcommand{\epro}{\end{proposition}}
\newcommand{\edfn}{\bbox\end{definition}}
\newcommand{\erem}{\bbox\end{remark}}
\newcommand{\exam}{\bbox\end{example}}
\newcommand{\ecor}{\end{corollary}}
\newcommand{\eprf}{\bbox\vspace{0.1in}}
\newcommand{\beqn}{\begin{equation}}
\newcommand{\eeqn}{\end{equation}}
 
\newcommand{\clm}{\begin{claim}}
\newcommand{\eclm}{\end{claim}}
 
\newcommand{\bbox}{\vrule height7pt width4pt depth1pt}

\newcommand{\sat}{\models}

\newcommand{\rimp}{\Rightarrow}
\newcommand{\union}{\cup}
\newcommand{\IN}{\mbox{$I\!\!N$}}

\renewcommand{\phi}{\varphi}
\newcommand{\Circ}{\mathbin{\mbox{{\small$\bigcirc$}}}}
 
\newcommand{\lt}{<}

\newcommand{\B}{{\cal B}}

\newcommand{\G}{{\cal G}}
\newcommand{\I}{{\cal I}}

\newcommand{\R}{{\cal R}}
\newcommand{\stc}{\, : \,} %colon with space around it

\renewcommand{\>}{\rangle}

\newcommand{\cf}{cf.~}

\newcommand{\ol}{\setlength{\itemsep}{0pt}\begin{enumerate}}
\newcommand{\eol}{\end{enumerate}\setlength{\itemsep}{-\parsep}}
\newcommand{\ul}{\setlength{\itemsep}{0pt}\begin{itemize}}
\newcommand{\dl}{\setlength{\itemsep}{0pt}\begin{description}}
\newcommand{\edl}{\end{description}\setlength{\itemsep}{-\parsep}}
\newcommand{\eul}{\end{itemize}\setlength{\itemsep}{-\parsep}}

\newcommand{\cI}{{\cal I}}

\newcommand{\cR}{{\cal R}}

\newcommand{\commentout}[1]{}

\newcommand{\bi}{\begin{itemize}}
\newcommand{\ei}{\end{itemize}}
\newcommand{\be}{\begin{enumerate}}
\newcommand{\ee}{\end{enumerate}}
 
\def\rarrowr{\buildrel{\smash{\raise 0.5ex \hbox{$\scriptstyle r$}}} \over
           {\smash{\mathop{\hbox to 1.3em {\rightarrowfill}}}} }

\newcommand{\ack}{\mbox{\it ack}}
\newcommand{\Gz}{\G_0}

\newcommand{\Isys}{{\bf I}^{{\it rep}}}
\newcommand{\sendbit}{{\sf sendbit}}

\newcommand{\recack}{{\it recack\/}}
\newcommand{\recbit}{{\it recbit\/}}

\newcommand{\bit}{{\it bit\/}}
\newcommand{\Pgkb}{{\sf TELL}}

\newcommand{\Pgbt}{{\sf BT}}
\newcommand{\sfa}{{\sf a}}

\newcommand{\gammafair}%
{\gamma^{{\it bt}}_{\mbox{\scriptsize{{\it fair}}}}}
\newcommand{\gammafairk}%
{\gamma^{{\it bt}}_{\mbox{\scriptsize{{\it fair,k}}}}}

\newcommand{\ACT}{{\it ACT\/}}
\newcommand{\Pg}{{\sf Pg}}

\newcommand{\cG}{{\cal G}}

\newcommand{\next}{\bigcirc}

\newcommand{\clock}{{\it clock}}

\newcommand{\ells}{\ell}
\newcommand{\Ifmp}{\mbox{$\I^{\kern 0.1ex \it fm'}$}}

\newtheorem{EXERCISE}{}[chapter]
{\end{EXERCISE}}
\newtheorem{HARDEX}[EXERCISE]{\llap{*}}
{\end{HARDEX}}
\newtheorem{SUPERHARD}[EXERCISE]{\llap{**}}
{\end{SUPERHARD}}
\newcommand{\oldindex}[1]{}

\newcommand{\glosipgg}{$\Isys(\Pg,\gamma,\pi)$}

\newcommand{\glospgpi}{$\Pg^\pi$}